\documentclass[11pt,a4paper,english]{amsart}
\usepackage[utf8]{inputenc}
\usepackage[english]{babel} 

\usepackage{amsmath} 
\usepackage{amsthm} 
\usepackage{graphicx} 
\usepackage{xcolor} 
\usepackage{amsfonts}
\usepackage[biblabel]{cite}
\usepackage{bm} 
\usepackage[hidelinks]{hyperref} 
\usepackage{amssymb} 
\usepackage{tikz-cd} 
\usepackage{amscd}
\usepackage{etoolbox}
\patchcmd{\thmhead}{(#3)}{#3}{}{}

\DeclareMathOperator{\ini}{in} 
\DeclareMathOperator{\ev}{ev} 

\DeclareMathOperator{\PRM}{PRM}
\DeclareMathOperator{\RM}{RM}

\DeclareMathOperator{\wt}{wt}

\newcommand{\F}{{\mathbb{F}}}
\newcommand{\fq}{\mathbb{F}_q}

\newcommand{\PP}{{\mathbb{P}}}
\newcommand{\PM}{{\mathbb{P}^{m}}}

\newcommand{\sd}{S_{d_1}/I(P^2)}
\newcommand{\sdd}{S_{d_2}/I(P^2)}
\newcommand{\sdh}{S_{d}/I(P^2)}
\newcommand{\sddh}{S_{d^\perp}^q/I(P^2)}
\newcommand{\adq}{(A^{d^\perp})^q}
\newcommand{\adql}{((A^{d^\perp})^q)^{(l)}}

\usepackage{mathtools} 
\DeclarePairedDelimiter\abs{\lvert}{\rvert}%
\DeclarePairedDelimiter\norm{\lVert}{\rVert}%

\makeatletter
\let\oldabs\abs
\def\abs{\@ifstar{\oldabs}{\oldabs*}}
\let\oldnorm\norm
\def\norm{\@ifstar{\oldnorm}{\oldnorm*}}
\makeatother

\newtheorem{thm}{Theorem}[section]
\newtheorem{prop}[thm]{Proposition}
\newtheorem{cor}[thm]{Corollary}
\newtheorem{lem}[thm]{Lemma}
\theoremstyle{definition}
\newtheorem{defn}[thm]{Definition} 

\newtheorem{rem}[thm]{Remark} 
 
\newtheorem{ex}[thm]{Example}

\title[Hulls of projective Reed-Muller codes over the projective plane]{Hulls of projective Reed-Muller codes over the projective plane}
\author{Diego Ruano and Rodrigo San-José}
\curraddr{
\texttt{Diego Ruano, Rodrigo San-José:} IMUVA-Mathematics Research Institute, Universidad de Valladolid, 47011 Valladolid (Spain).
}
\email{diego.ruano@uva.es; rodrigo.san-jose@uva.es}

\thanks{This work was supported in part by the following grants: Grant TED2021-130358B-I00 funded by MCIN/AEI/10.13039/501100011033 and by the ``European Union NextGenerationEU/PRTR'', by Grants PID2022-138906NB-C21 and PID2022-138906NB-C21 funded by MICIU/AEI/10.13039/501100011033 and by ERDF/EU, QCAYLE project funded by MCIN, the European Union NextGenerationEU (PRTR C17.I1) and Junta de Castilla y Le\'on, and FPU20/01311 funded by the Spanish Ministry of Universities.}



\begin{document}

\maketitle

\begin{abstract}
By solving a problem regarding polynomials in a quotient ring, we obtain the relative hull and the Hermitian hull of projective Reed-Muller codes over the projective plane. The dimension of the hull determines the minimum number of maximally entangled pairs required for the corresponding entanglement-assisted quantum error-correcting code. Hence, by computing the dimension of the hull we now have all the parameters of the symmetric and asymmetric entanglement-assisted quantum error-correcting codes constructed with projective Reed-Muller codes over the projective plane. As a byproduct, we also compute the dimension of the Hermitian hull for affine Reed-Muller codes in 2 variables. 
\end{abstract}

\section{Introduction}
Evaluation codes, have a rich algebraic structure and can be studied using tools from commutative algebra  \cite{jaramillo,hiramdualevaluationcode,villarrealminimumdistancefunctions}. In particular, projective Reed-Muller codes are a family of evaluation codes obtained by evaluating homogeneous polynomials of a given degree at the projective space \cite{lachaud,sorensen}. In \cite{lachaud} it is shown that these codes can outperform affine Reed-Muller codes in some instances. Taking into account that Reed-Muller codes were one of the first families of linear codes used to construct quantum error-correcting codes (QECCs) \cite{steaneRM}, it is natural to consider projective Reed-Muller codes for constructing quantum codes. When one evaluates over the projective line, this family corresponds to projective Reed-Solomon codes, which have been used for constructing QECCs in various contexts \cite{ball,sanjoseSSCPRS}. As we will see, by using evaluation codes we translate problems about quantum and classical codes to questions regarding polynomials in a quotient ring. 

The importance of QECCs is growing in parallel to the interest in quantum computing, as QECCs are necessary to achieve fault tolerant computation \cite{shorfaulttolerant}. One can construct QECCs from classical linear codes using the CSS construction and the Hermitian construction \cite{css1,css2,kkks}, but these constructions require to have some self-orthogonality conditions for the corresponding classical codes. Using entanglement assistance, it is possible to remove the self-orthogonality restrictions, giving rise to entanglement-assisted quantum error-correcting codes (EAQECCs) \cite{brun}. Both the CSS construction and the Hermitian construction can be generalized to this context, see Theorems \ref{asimetricos} and \ref{hermitica} from \cite{galindoasymmetric} and  \cite{galindoentanglement}, respectively.

For the CSS construction, one can consider two codes $C_1$ and $C_2$, and the minimum number required of maximally entangled quantum states is equal to $c:=\dim C_1-\dim C_1\cap C_2^\perp$. Therefore, the dimension of the \textit{relative hull of $C_1$ with respect to $C_2$}, defined as
$$
\text{Hull}_{C_2}(C_1):=C_1\cap C_2^\perp,
$$
determines the parameter $c$ \cite{relativehull}. For the Hermitian construction, we only use one code $C$, and the parameter $c$ is given by $\dim C-\dim C\cap C^{\perp_h}$, where $C^{\perp_h}$ is the Hermitian dual of $C$. The \textit{Hermitian hull} of $C$ is thus defined as 
$$
\text{Hull}^H(C):=C\cap C^{\perp_h}.
$$

Going back to projective Reed-Muller codes, as they are evaluation codes, we can view their codewords as classes of polynomials in a quotient ring. This motivates Section \ref{sechull} of this paper, where we compute bases of polynomials for some appropriate subspaces of the quotient rings associated to projective Reed-Muller codes by using Gröbner bases techniques. As a consequence of this computation, we give a basis for the relative hull of projective Reed-Muller codes over the projective plane $\mathbb{P}^2$. We also estimate the dimension of the Hermitian hull, and give bases in some cases. The estimate that we obtain is sharp in all the cases we have checked. As a byproduct of these computations, we also compute the Hermitian hull of affine Reed-Muller codes in 2 variables, which was known to be trivial in some cases \cite{galindostabilizer,sarvepallinonbinaryquantumRM}, but was not known in general. With this knowledge, in Section \ref{secquant} we give the parameters of the EAQECCs obtained by using projective Reed-Muller codes over $\mathbb{P}^2$, since the dimension of the hull determines the parameter $c$. Although projective Reed-Muller codes had already been used to construct QECCs in some particular cases \cite{libroprm}, the cases that required entanglement assistance had not been yet addressed. We focus on the case of $\mathbb{P}^2$ because it offers a good trade-off between providing long codes over a small finite field and avoiding computations that are too involved, making explicit formulas unfeasible. For the general case of $\mathbb{P}^m$, one has to make additional assumptions, such as restricting to the Euclidean case and requiring $C_1=C_2$, see \cite{kaplanHullsPRM}.

\section{Preliminaries}

We consider the finite field $\fq$ with $q$ elements, and the projective space $\PM$ over $\F_{q}$. Throughout this work, we will fix representatives for the points of $\PM$: for each point $[Q]\in \PM$, we choose the representative whose leftmost entry is equal to 1. We will denote by $P^m=\{Q_1,\dots,Q_n\}$, with $n=\abs{P^m}=\frac{q^{m+1}-1}{q-1}$, the set of representatives that we have chosen (seen as points in the affine space $\mathbb{A}^{m+1}$). For a set of points $A\subset \mathbb{A}^{m+1}$ we will denote by $[A]$ the set of points $\{[a_0:\cdots:a_m] \mid (a_0,\dots,a_m)\in A\setminus\{0\}\}$ (using the representatives that we have chosen).

We consider now the polynomial ring $S=\fq [x_0,\dots,x_m]$. The \textit{evaluation map} is the $\F_q$-linear map defined by
$$
\ev:S \rightarrow \fq^{n},\:\: f\mapsto \left(f(Q_1),\dots,f(Q_n)\right).
$$
Let $d$ be a positive integer. If we consider $S_d\subset S$, the set of homogeneous polynomials of degree $d$, we have that $\ev(S_d)$ is the \textit{projective Reed-Muller code} of degree $d$, which we will denote by $\PRM_d(q,m)$, or $\PRM_d(m)$ if there is no confusion about the field. For a code $C\subset \F_q^n$, we denote its minimum distance by $\wt(C)$. The following results about the parameters of projective Reed-Muller codes and their duality appear in \cite{sorensen}.

\begin{thm}\label{paramPRM}
The projective Reed-Muller code $\PRM_d(q,m)$, $1\leq d\leq m(q-1)$, is an $[n,k]$-code with 
$$
\begin{aligned}
&n=\frac{q^{m+1}-1}{q-1},\\
&k=\sum_{t\equiv d\bmod q-1,0<t\leq d}\left( \sum_{j=0}^{m+1}(-1)^j\binom{m+1}{j}\binom{t-jq+m}{t-jq}   \right).\\
\end{aligned}
$$
For the minimum distance, we have
$$
\wt(\PRM_d(q,m))=(q-s)q^{m-r-1}, \text{ where } \;
d-1=r(q-1)+s, \;0\leq s <q-1.
$$
\end{thm}
\begin{thm}\label{dualPRM}
Let $1\leq d\leq m(q-1)$ and let $d^\perp=m(q-1)-d$. Then
$$
\begin{aligned}
&\PRM_d^\perp(q,m)=\PRM_{d^\perp}(q,m) &\text{ if } d\not\equiv 0\bmod q-1, \\
&\PRM_d^\perp(q,m)=\PRM_{d^\perp}(q,m)+\langle (1,\dots,1) \rangle &\text{ if } d\equiv 0\bmod q-1.
\end{aligned}
$$
\end{thm}

\begin{rem}
Theorem \ref{dualPRM} states that, if $d\not\equiv 0 \bmod q-1$ the dual of a projective Reed-Muller code is another projective Reed-Muller code. If we define $\PRM_0(2)=\langle (1,\dots,1)\rangle$, then for $d=m(q-1)$ we can also say that the dual of a projective Reed-Muller code is another projective Reed-Muller code. Hence, for $m=2$, the case we are going to study in this paper, the only case in which the dual of a projective Reed-Muller code is not another projective Reed-Muller code is when $d=q-1$. 
\end{rem}

With respect to affine Reed-Muller codes, we denote them by $\RM_d(q,m)$, or by $\RM_d(m)$ if there is no confusion about the field. We have the following results about their parameters and their duality from \cite{delsarteRM,kasamiRM}.

\begin{thm}\label{paramRM}
The Reed-Muller code $\RM_d(q,m)$, $0\leq d\leq m(q-1)$, is an $[n,k]$-code with 
$$
\begin{aligned}
&n=q^m,\\
&k=\sum_{t=0}^d \sum_{j=0}^m (-1)^j\binom{m}{j}\binom{t-jq+m-1}{t-jq}.\\
\end{aligned}
$$
For the minimum distance, we have
$$
\wt(\RM_d(q,m))=(q-s)q^{m-r-1}, \text{ where } \;
d=r(q-1)+s, \;0\leq s <q-1.
$$
\end{thm}

\begin{thm}\label{dualRM}
Let $0\leq d\leq m(q-1)$. Then
$$
\RM_d^\perp(q,m)=\RM_{m(q-1)-d-1}(q,m).
$$
\end{thm}

Going back to projective Reed-Muller codes, we have seen that $\PRM_d(m)=\ev(S_d)$, which gives the isomorphism $$\PRM_d(m)\cong S_d/(I(P^m)\cap S_d)\cong (S_d+I(P^m))/I(P^m),$$ where $I(P^m)$ is the vanishing ideal of $P^m$. This is because, if we restrict $\ev$ to $S_d$, the polynomials in the kernel are precisely the polynomials from $S_d$ that vanish at each of the points of $P^m$, which are the polynomials in $I(P^m)\cap S_d$. This isomorphism allows us to view the vectors of the code as polynomials in a quotient ring. It is important to note that two polynomials in $S/I(P^m)$ have the same evaluation if and only if their classes in $S/I(P^m)$ are the same. This is why we can discuss linear independence both in $\PRM_d(m)\subset \fq^n$ and in $S/I(P^m)$.

Moreover, we can express many important aspects of the code purely in terms of polynomials, for example the minimum distance \cite{jaramillo} or their duals \cite{hiramdualevaluationcode}. The theory of Gröbner bases is one of the main tools that are used for studying evaluation codes using this approach. In the rest of this section, we introduce some of the Gröbner-related results for projective Reed-Muller codes that we will use in the rest of the paper. 

In what follows, we will abuse the notation and denote $S_d/I(P^m)=(S_d+I(P^m))/I(P^m)$. Moreover, for a polynomial $f\in S$, we will use the same notation $f$ for both the polynomial and its class in $S/I(P^m)$. We refer the reader to \cite{cox} for the basic concepts of Gröbner bases. For $f\in S$, we denote by $\ini(f)$ the leading monomial of $f$ (without the coefficient). For an ideal $I\subset S$, $\ini(I)$ denotes the ideal generated by the leading monomials of the polynomials in $I$. We have the following result for the vanishing ideal of $P^m$ from \cite{sanjoseSSCPRM}.

\begin{thm}\label{vanishingideal}
The vanishing ideal of $P^m$ is generated by the following polynomials:
$$
\begin{aligned}
I(P^m)=&\langle  x_0^2-x_0,x_1^{q}-x_1,x_2^{q}-x_2,\dots,x_m^{q}-x_m,(x_0-1)(x_1^2-x_1),\\
&(x_0-1)(x_1-1)(x_2^2-x_2),\dots,(x_0-1)\cdots(x^2_{m-1}-x_{m-1}),(x_0-1)\cdots(x_m-1) \rangle.
\end{aligned}
$$
Moreover, these generators form a universal Gröbner basis of the ideal $I(P^m)$ (i.e., they form a Gröbner basis for any monomial order), and we have that
$$
\ini(I(P^m))=\langle x_0^2,x_1^{q},x_2^{q},\dots,x_m^{q},x_0x_1^2,x_0x_1x_2^2,\dots,x_0x_1\cdots x_{m-1}^2,x_0x_1\cdots x_m \rangle .
$$
\end{thm}

In this work we will study the case $m=2$, in which we have
$$
I(P^2)=\langle x_0^2-x_0,x_1^q-x_1,x_2^q-x_2,(x_0-1)(x_1^2-x_1),(x_0-1)(x_1-1)(x_2-1) \rangle.
$$

We introduce now the bases of polynomials that we will use in the following sections.

\begin{lem}\label{basehomogenea}
Let $1\leq d \leq 2(q-1)$. We consider the following sets of monomials:
$$
\begin{aligned}
&A_1^d=\{ x_0^{a_0}x_1^{a_1}x_2^{a_2}\mid a_0>0, a_0+a_1+a_2=d, 0\leq a_1,a_2\leq q-1 \},\\
&A_2^d=\{ x_1^{a_1}x_2^{a_2}\mid a_1>0,a_1+a_2=d, 0\leq a_2\leq q-1\} ,\\
&A_3^d=\{ x_2^{d} \}.
\end{aligned}
$$
Then, $A^d=A_1^d\cup A_2^d\cup A_3^d$ forms a basis for $S_d/I(P^2)$.
\end{lem}
\begin{proof}
$A^d$ is a basis for $S_d/I(\PP^2)_d$ (for example, see \cite{projectivefootprint}), where $I(\PP^2)$ is the vanishing ideal of $\PP^2$, i.e, the ideal generated by the homogeneous polynomials that vanish in all the points of $\PP^2$. Therefore, the image by the evaluation map of $A^d$ is a basis for $\PRM_d(2)$, which means that the classes of these polynomials in $S/I(P^2)$ are also a basis for $S_d/I(P^2)$.
\end{proof}

\begin{rem}\label{remdim}
Let $1\leq d\leq 2(q-1)$ and $k=\dim \RM_{d-1}(2)$. Then it is clear that $\abs{A_1^d}=k$, and the previous result shows that we have the following formula:
$$
\dim \PRM_{d}(2)=\begin{cases}
    k+d+1 &\text{ if } 1\leq d\leq q-1, \\
    k+q+1 &\text{ if } q\leq d< 2(q-1).
\end{cases}
$$
\end{rem}

From \cite{sanjoseSSCPRM}, we have the following lemma about $S/I(P^2)$.

\begin{lem}\label{basePm}
The set of monomials $\{x_1^{a_1}x_2^{a_2}, x_0x_2^{a_2},x_0x_1 \mid 0\leq a_i\leq q-1,1\leq i\leq 2\}$ is a basis for $S/I(P^2)$.
\end{lem}

We show now how to express any monomial from $A^d$ in terms of the basis from Lemma \ref{basePm}. The following definition is useful for this purpose.

\begin{defn}\label{D:overline}
For an integer $z\geq 0$, we denote by $\overline{z}$ the integer $1\leq \overline{z}\leq q-1$ such that $\overline{z}\equiv z\bmod q-1$ if $z>0$, and $\overline{z}=0$ if $z=0$. 
\end{defn}

\begin{rem}\label{R:x0}
Any monomial $x_0^{\alpha_0}x_1^{\alpha_1}x_2^{\alpha_2}$ with $\alpha_0>0$ is equivalent to $x_0^{\alpha'_0}x_1^{\alpha_1}x_2^{\alpha_2}$ in $S/I(P^2)$, for any $\alpha'_0>0$, because we have the polynomial $x_0^2-x_0$ in $I(P^2)$. In particular, $x_0^{\alpha_0}x_1^{\alpha_1}x_2^{\alpha_2}$ is equivalent to $x_0x_1^{\alpha_1}x_2^{\alpha_2}$ if $\alpha_0>0$. Moreover, in Definition \ref{D:overline} we treat the case $z=0$ separately so that we have
$$
x_0^{a_0}x_1^{a_1}x_2^{a_2}\equiv x_0^{\overline{a_0}}x_1^{\overline{a_1}}x_2^{\overline{a_2}} \bmod I(P^2),
$$
for any $0\leq a_0,a_1,a_2\leq 2(q-1)$. Notice that, for evaluation codes, having exponent $q-1$ is not the same as $0$ since, for instance,
$$
x_0^0=1\not\equiv x_0^{q-1}\bmod I(P^2).
$$
This can be checked by evaluating $1$ and $x_0^{q-1}$ at any point of the form $(0,1,x_2)$, $x_2\in \fq$ (recall that two polynomials are equivalent modulo $I(P^2)$ if and only if they have the same evaluation). 
\end{rem}

The following Lemma from \cite{sanjoseSSCPRM} shows how to express any monomial in terms of the basis from Lemma \ref{basePm}. 

\begin{lem}\label{divisionP2}
Let $a_0,a_1,a_2$ be integers. 
\begin{enumerate}
    \item If $a_0=0$, then
    $$
    x_1^{a_1}x_2^{a_2}\equiv x_1^{\overline{a_1}}x_2^{\overline{a_2}} \bmod I(P^2).
    $$
    \item If $a_1=0$, then
    $$
    x_0^{a_0}x_2^{a_2}\equiv x_0x_2^{\overline{a_2}} \bmod I(P^2).
    $$
    \item If $a_0>0$ and $a_1>0$, then 
    $$
    \begin{aligned}
    x_0^{a_0}x_1^{a_1}x_2^{a_2}&\equiv x_1^{\overline{a_1}}x_2^{\overline{a_2}}+x_0x_2^{\overline{a_2}}-x_2^{\overline{a_2}}+x_0x_1-x_0-x_1+1 \bmod I(P^2)\\
    &\equiv x_1^{\overline{a_1}}x_2^{\overline{a_2}}+(x_0-1)(x_2^{\overline{a_2}}+x_1-1) \bmod I(P^2).
    \end{aligned}
    $$
\end{enumerate}
\end{lem}

In particular, Lemma \ref{divisionP2} allows us to express all the monomials from the basis in Lemma \ref{basehomogenea} in terms of the basis from Lemma \ref{basePm}. This is crucial because the idea of the next section is, for $1\leq d_1 \leq d_2 \leq 2(q-1)$, to consider the bases from Lemma \ref{basehomogenea} for $S_{d_1}/I(P^2)$ and $S_{d_2}/I(P^2)$, and express them in terms of the basis for $S/I(P^2)$ from Lemma \ref{basePm} using Lemma \ref{divisionP2}. Once we have all the polynomials expressed in this way, it is easier to find the polynomials lying in $\sd\cap \sdd$, which, as we will see, determines the relative hull of $\PRM_{d_1}(2)$ and $\PRM_{d_2}(2)$. For the Hermitian case, the ideas are also very similar, although the computations are more involved. 

\section{Computing the hulls of projective Reed-Muller codes}\label{sechull}
The aim of this section is to obtain bases for the relative hull and Hermitian hull of projective Reed-Muller codes. We do this by computing first a basis of polynomials for some appropriate subspaces of $S/I(P^2)$. We deduce, in particular, the dimension of the hull, which will determine the entanglement parameter $c$ for the EAQECCs constructed with projective Reed-Muller codes in Section \ref{secquant}.

\subsection{Euclidean hull}\label{seceu}
In this subsection we compute a basis for $\text{Hull}_{C_2}(C_1)=C_1\cap C_2^\perp$, the relative hull of $C_1$ and $C_2$, when $C_i=\PRM_{d_i}(2)$, for $i=1,2$. For $\PRM_{d_2}^\perp(2)$, by Theorem \ref{dualPRM} we have that, if $d_2\not= q-1$, then $\PRM_{d_2}^\perp(2)=\PRM_{d_2^\perp}(2)$, where $d_2^\perp=2(q-1)-d_2$ (assuming $\PRM_0(2)=\langle (1,\dots,1)\rangle$). We avoid the case where $d_2=q-1$ because, by Theorem \ref{dualPRM}, in that case we have $\PRM_{d_2}^\perp(2)=\PRM_{d_2^\perp}(2)+\langle (1,\dots,1)\rangle$, which is no longer isomorphic to $S_{d_2^\perp}/I(P^2)$. Assuming $d_2\not=q-1$, the problem of obtaining a basis for $\PRM_{d_1}(2)\cap\PRM_{d_2}^\perp(2)$ becomes equivalent to computing a basis for $\PRM_{d_1}(2)\cap\PRM_{d_2}(2)$, for any $1\leq d_1\leq d_2\leq 2(q-1)$, $d_2\not = q-1$, and that is the problem we solve in what follows. In \cite[Thm. 10.7]{libroprm} the authors construct quantum codes using pairs of projective Reed-Muller codes such that the dual of one of the codes is contained in the other. In particular, they obtain the following result.

\begin{lem}\label{hullfacilprm}
Let $1\leq d_1\leq d_2 \leq m(q-1)$. If $d_1\equiv d_2\bmod q-1$, then $\PRM_{d_1}(m)\subset \PRM_{d_2}(m)$. 
\end{lem}

Therefore, when $d_1\equiv d_2 \bmod q-1$, with $d_2\not= q-1$, the relative hull is straightforward to obtain. To avoid making exceptions in many of the following results, we exclude the case $d_1\equiv d_2\bmod q-1$, which is already covered by Lemma \ref{hullfacilprm}.

Because of the isomorphism $S_{d_i}/I(P^2)\cong \PRM_{d_i}(2)$, for $i=1,2$, the problem of computing a basis for $\PRM_{d_1}(2)\cap \PRM_{d_2}(2)$ can be understood as computing a basis for $ S_{d_1}/I(P^2)\cap S_{d_2}/I(P^2)$, as a subspace of $S/I(P^2)$. Hence, computing the dimension of $ S_{d_1}/I(P^2)\cap S_{d_2}/I(P^2)$ for arbitrary $1\leq d_1\leq d_2\leq 2(q-1)$ gives the parameter $c$ for any pair of projective Reed-Muller codes over $P^2$ (except when we have $d_2=q-1$). We give now some preliminary results. 

\begin{lem}\label{noesta1}
We have that $1\not \in S_{d}/I(P^2)$ for $1\leq d \leq 2(q-1)$. 
\end{lem}
\begin{proof}
If $d\leq q-1$ and we had $1\in S_{d}/I(P^2)$, then we would have the evaluation of $x_0^d-1$ in $\PRM_d(2)$, which has Hamming weight $q^2+q+1-q^2=q+1<(q-d+1)q=\wt(\PRM_d(2))$. For $q\leq d\leq 2(q-1)$, if we had $1\in S_{d}/I(P^2)$, then we would also have the evaluation of $x_0^d-x_0^{q-1}x_1^{d-(q-1)}+x_1^d-1\equiv x_0^d+(1-x_0)x_1^d-1\bmod I(P^2)$ in $\PRM_d(2)$ (recall Remark \ref{R:x0}), which has Hamming weight $1<\wt(\PRM_d(2))$. 
\end{proof}

Let $f\in S_{d_1}$. The following lemmas show when we have $f\in S_{d_2}/I(P^2)$ depending on the monomials that are in the support of $f$. This will allow us to determine which polynomials are in $\sd\cap\sdd$ and, thus, to obtain a basis.

\begin{lem}\label{lemaA1}
Let $1\leq d_1<d_2\leq 2(q-1)$. We have that the classes of the monomials in $A_1^{d_1}$ are contained in $ S_{d_2}/I(P^2)$.
\end{lem}
\begin{proof}
Let $ x_0^{a_0}x_1^{a_1}x_2^{a_2}\in A_1^{d_1}$. Then $ x_0^{a_0+d_2-d_1}x_1^{a_1}x_2^{a_2}\in \sdd$, and  $ x_0^{a_0}x_1^{a_1}x_2^{a_2}\equiv x_0^{a_0+d_2-d_1}x_1^{a_1}x_2^{a_2}\bmod I(P^2)$ by Remark \ref{R:x0}.
\end{proof}

Let $1\leq d_1<d_2\leq 2(q-1)$. We note that the monomials in $A_1^{d_1}$ and $A_1^{d_2}$ generate $\RM_{d_1-1}(2)$ and $\RM_{d_2-1}(2)$, respectively, when considering their evaluation in $[\{1\}\times \fq^2]$. Thus, $\dim\left( S_{d_1}/I(P^2)\cap S_{d_2}/I(P^2) \right)\geq \dim \RM_{d_1-1}(2)$.

\begin{ex}\label{running1} 
Let $\fq=\F_4$, and we consider $d_1=4<5=d_2$. By Lemma \ref{lemaA1}, we have that $A_1^4$ is in $S_4/I(P^2)\cap S_5/I(P^2)$. In this case, we have 
$$
A_1^4=\{x_0^4,x_0^3x_1,x_0^3x_2,x_0^2x_1^2,x_0^2x_1x_2,x_0^2x_2^2,x_0x_1^3,x_0x_1^2x_2,x_0x_1x_2^2,x_0x_2^3\}.
$$
If we multiply all the elements of $A_1^4$ by $x_0$, we obtain a set of monomials of degree 5 which have the same evaluation at $P^2$ as these monomials (see Remark \ref{R:x0}). 
\end{ex}

\begin{lem}\label{lemaa2}
Let $1\leq d_1<d_2\leq 2(q-1)$ with $d_1\not\equiv d_2\bmod q-1$. Let $f\in S_{d_1}$ such that it only has monomials from $ A_2^{d_1}$ in its support, i.e., it can be expressed as
$$
f=\sum_{a_1+a_2=d_1,0<a_1,0\leq a_2\leq q-1}\lambda_{a_1,a_2} x_1^{a_1}x_2^{a_2}, \lambda_{a_1,a_2}\in \fq.
$$
Let $Y=\{0,1,\dots,\min\{d_1-1,d_2-q\}\}$ and $Y_f=\{0\leq a_2\leq q-1 \mid \lambda_{d_1-a_2,a_2}\neq 0\}$. Then $f\in \sdd$ if and only if $Y_f\subset Y$. 
\end{lem}
\begin{proof}
Assuming that $f\in \sdd$, there is also an expression
\begin{equation}\label{fa2}
f\equiv \sum_{x_0^{a_0}x_1^{a_1}x_2^{a_2}\in A^{d_2}} \gamma_{a_0,a_1,a_2}x_0^{a_0}x_1^{a_1}x_2^{a_2}  \bmod I(P^2), \gamma_{a_0,a_1,a_2}\in \fq.
\end{equation}
We have $f(0,0,1)=0$, which means that $\gamma_{0,0,d_2}=0$. We consider now a monomial order with $x_0<x_1<x_2$, and let $\ini(f)=x_1^{a_1}x_2^{a_2}\equiv x_1^{\overline{a_1}}x_2^{a_2}\bmod I(P^2)$, with $a_1>0$, $0\leq a_2\leq q-1$ and $a_1+a_2=d_1$. Therefore, because of (\ref{fa2}), we must have some monomial in $A^{d_2}$ such that its expression in the basis from Lemma \ref{basePm} contains $x_1^{\overline{a_1}}x_2^{a_2}$ in its support. The only monomials that satisfy that are $x_1^{\overline{a_1}+q-1}x_2^{a_2}$ if $d_2=\overline{a_1}+a_2+(q-1)$, or $x_0^{c_0}x_1^{\overline{a_1}}x_2^{a_2}$, with $c_0=d_2-\overline{a_1}-a_2>0$. In the first case, we have $d_1\equiv d_2\bmod q-1$, but we are assuming $d_1\not \equiv d_2\bmod q-1$. In the other case, by Lemma \ref{divisionP2} we have
$$
x_0^{c_0}x_1^{\overline{a_1}}x_2^{a_2}\equiv x_1^{\overline{a_1}}x_2^{a_2}+(x_0-1)(x_2^{a_2}+x_1-1)\bmod I(P^2).
$$
Hence, to obtain $x_1^{\overline{a_1}}x_2^{a_2}$ in the right-hand side of (\ref{fa2}), we must have $\gamma_{c_0,\overline{a_1},a_2}=\lambda_{a_1,a_2}$. We denote now $(A^{d_2})^{(1)}:=A^{d_2}\setminus \{ x_0^{c_0}x_1^{\overline{a_1}}x_2^{a_2}\}$. We obtain
\begin{equation}\label{exp1}
\begin{aligned}
 f^{(1)}=&f-\lambda_{a_1,a_2}x_1^{a_1}x_2^{a_2}\\
 \equiv& \sum_{x_0^{\alpha_0}x_1^{\alpha_1}x_2^{\alpha_2}\in (A^{d_2})^{(1)}} \gamma_{\alpha_0,\alpha_1,\alpha_2}x_0^{\alpha_0}x_1^{\alpha_1}x_2^{\alpha_2} +\gamma_{c_0,\overline{a_1},a_2}(x_0-1)(x_2^{a_2}+x_1-1) \bmod I(P^2).
\end{aligned}
\end{equation}
Now we have $\ini(f^{(1)})< \ini(f)$. We can consider $\ini(f^{(1)})=x_1^{b_1}x_2^{b_2}$ and argue as before to obtain a polynomial $f^{(2)}$ such that $\ini(f^{(2)})<\ini(f^{(1)})$, which can be expressed as in (\ref{exp1}) in terms of a set $(A^{d_2})^{(2)}=A^{d_2}\setminus \{ x_0^{c_0}x_1^{\overline{a_1}}x_2^{a_2},x_0^{c'_0}x_1^{\overline{b_1}}x_2^{b_2}\}$.

We can do this until we get $f^{(l)}=0$ for some $l\geq 0$. At that step, we have
\begin{equation}\label{exp2}
0\equiv \sum_{x_0^{\alpha_0}x_1^{\alpha_1}x_2^{\alpha_2}\in (A^{d_2})^{(l)}} \gamma_{\alpha_0,\alpha_1,\alpha_2}x_0^{\alpha_0}x_1^{\alpha_1}x_2^{\alpha_2} +(x_0-1)\sum_{a_2\in Y_f}\gamma_{c_0,\overline{a_1},a_2}(x_2^{a_2}+x_1-1) \bmod I(P^2),
\end{equation}
where $Y_f=\{0\leq a_2\leq q-1 \mid \lambda_{d_1-a_2,a_2}\neq 0\}$. With this notation, we have that $(A^{d_2})^{(l)}=A^{d_2}\setminus \bigcup_{a_2\in Y_f}\{ x_0^{d_2-(\overline{d_1-a_2})-a_2}x_1^{\overline{d_1-a_2}}x_2^{a_2}\}$. If we express all the monomials in (\ref{exp2}) in terms of the basis from Lemma \ref{basePm}, then we must have the coefficient of each element of the basis equal to 0. The monomials from (\ref{exp2}) in the second sum are already expressed in terms of the basis from Lemma \ref{basePm}. If we focus on the monomial $x_2^{a_2}$ for some $a_2\in Y_f$ with $a_2>0$, we see that all the monomials $x_0^{\alpha_0}x_1^{\alpha_1}x_2^{a_2}\in (A^{d_2})^{(l)}$ with $0<\alpha_0$, $0<\alpha_1\leq q-1$, $\alpha_1\neq \overline{d_1-a_2}$, and $\alpha_0+\alpha_1+a_2=d_2$, have $x_2^{a_2}$ in their expression in terms of the basis from Lemma \ref{basePm} by Lemma \ref{divisionP2}:
\begin{equation}\label{expnueva}
x_0^{\alpha_0}x_1^{\alpha_1}x_2^{a_2}\equiv x_1^{\alpha_1}x_2^{a_2}+(x_0-1)(x_2^{a_2}+x_1-1) \bmod I(P^2).
\end{equation}
In fact, these are the only monomials from $(A^{d_2})^{(l)}$ with $x_2^{a_2}$ in their expression (note that the monomial with $\alpha_1=\overline{d_1-a_2}$ is not in $(A^{d_2})^{(l)}$). However, if we have $\gamma_{\alpha_0,\alpha_1,a_2}\neq 0$, some other monomial from $(A^{d_2})^{(l)}$ must cancel the monomial $x_1^{\alpha_1}x_2^{a_2}$ that appears in (\ref{expnueva}) from (\ref{exp2}). The only other monomial in $(A^{d_2})^{(l)}$ with $x_1^{\alpha_1}x_2^{a_2}$ in its support when expressed in terms of the basis from Lemma \ref{basePm} is $x_1^{\alpha_1+q-1}x_2^{a_2}$, if $\alpha_1+a_2+q-1=d_2$ (which implies $\alpha_0=q-1$). 

Therefore, if $d_2\leq q-1$, given $a_2\in Y_f$, $a_2>0$, the monomial $x_2^{a_2}$ from (\ref{exp2}) cannot be cancelled with any monomial from $(A^{d_2})^{(l)}$, which means that we must have $\gamma_{c_0,\overline{a_1},a_2}=\lambda_{a_1,a_2}=0$, a contradiction with the fact that $a_2\in Y_f$. This means that there is no $a_2\in Y_f$ with $a_2>0$. Taking this into account, the only possible term in the second sum of (\ref{exp2}) corresponds to the case $a_2=0$, and we have $\gamma_{c_0,\overline{a_1},a_2}(x_0-1)(x^{a_2}+x_1-1)=\gamma_{c_0,\overline{d_1},0}(x_0-1)x_1$. This polynomial cannot be generated by polynomials from $A^{d_2}$ because its evaluation has Hamming weight $q^2+q+1-q^2-1=q<\wt(\PRM_{d_2}(2))$ if $d_2\leq q-1$. Thus, if $d_2\leq q-1$, we have must have $Y_f=\emptyset$ and $f=0$. 

Lets assume now that $d_2\geq q$. For each $a_2\in Y_f$ with $a_2 < \overline{d_2}$, we can consider $\alpha_1=\overline{d_2}-a_2>0$. We have seen that
$$
x_0^{\alpha_0}x_1^{\alpha_1}x_2^{a_2}-x_1^{\alpha_1+q-1}x_2^{a_2}\equiv (x_0-1)(x_2^{a_2}+x_1-1) \bmod I(P^2).
$$
Note that in the monomials that we have excluded to obtain $(A^{d_2})^{(l)}$ from $A^{d_2}$, we have that the exponent of $x_1$ is equivalent to $d_1-a_2$ modulo $q-1$. The $\alpha_1$ that we have chosen in this case is equivalent to $d_2-a_2$ modulo $q-1$, which means that the corresponding monomial is still in $(A^{d_2})^{(l)}$, unless $d_1\equiv d_2 \bmod q-1$, which is the case that we do not cover. Hence, for every $a_2\in Y_f$ with $a_2<\overline{d_2}$, if we choose $\gamma_{\alpha_0,\alpha_1,a_2}=-\gamma_{0,\alpha_1+q-1,a_2}=\lambda_{d_1-a_2,a_2}$, the polynomial  $(x_0-1)(x_2^{a_2}+x_1-1)$ is cancelled from (\ref{exp2}). If we had some $a_2\in Y_f$ with $a_2\geq \overline{d_2}$, we can argue as in the previous case and obtain that $\lambda_{d_1-a_2,a_2}=0$, a contradiction. 

We also have that $a_2\leq d_1-1$ for every $a_2\in Y_f$. Therefore, $Y_f\subset \{0,1,\dots,\min\{d_1-1,{\overline{d_2}}-1\}\}$. Thus, $Y_f\subset Y=\{0,1,\dots,\min\{d_1-1,d_2-q\}\}$, where if $d_2-q<0$ we understand that $Y=\emptyset$, which covers the case with $d_2\geq q$ and the case with $d_2\leq q-1$. 

On the other hand, let $f\in S_{d_1}$ such that it only has monomials from $A_2^{d_1}$ in its support, and with $Y_f\subset Y$. For each $a_2\in Y_f$ we have
$$
x_1^{d_1-a_2}x_2^{a_2}\equiv x_1^{d_2-a_2}x_2^{a_2}-x_0^{d_2-\overline{d_2}}x_1^{\overline{d_2}-a_2}x_2^{a_2}+x_0^{d_2-d_1}x_1^{d_1-a_2}x_2^{a_2} \bmod I(P^2).
$$
This is easy to check because both sides have the same evaluation at $P^2=[\{1\}\times \fq^2]\cup[\{0\}\times \{1\}\times \fq]\cup\{[0:0:1]\}$. Hence, $x_1^{d_1-a_2}x_2^{a_2}\in \sd\cap \sdd$ for each $a_2\in Y_f$, which means that $f\in \sd\cap \sdd$. 
\end{proof}

\begin{ex}\label{running2}
We continue with Example \ref{running1}. Using the notation from Lemma \ref{lemaa2}, we have $Y=\{0,1\}$ and we obtain that the set of monomials $\{x_1^4,x_1^3x_2\}$ is contained in $S_4/I(P^2)\cap S_5/I(P^2)$. In fact, following the proof of Lemma \ref{lemaa2}, we have that
$$
x_1^4\equiv x_1^5-x_0^3x_1^2+x_0x_1^4 \bmod I(P^2),
$$
$$
x_1^3x_2\equiv x_1^4x_2-x_0^3x_1x_2+x_0x_1^3x_2\bmod I(P^2).
$$
These equivalences can be easily checked by evaluating both sides at $P^2=[\{1\}\times \F_4^2]\cup[\{0\}\times \{1\}\times \F_4]\cup\{[0:0:1]\}$.
\end{ex}

\begin{lem}\label{lemaa3}
Let $1\leq d_1<d_2\leq 2(q-1)$ such that $d_1\not\equiv d_2\bmod q-1$. 
There is some $f\in S_{d_1}$ with $x_2^{d_1}$ in its support and such that $f\in \sdd$ if and only if $d_1\geq q$.
\end{lem}
\begin{proof}
Let $f\in S_{d_1}$. By Lemma \ref{lemaA1}, if there is some monomial from $A_1^{d_1}$ in the support of $f$, we can consider the polynomial $f'$ obtained by subtracting that monomial from $f$, and $f'\in \sdd$ if and only if $f\in \sdd$. Therefore, we can assume that the support of $f$ is contained in $ A_2^{d_1}\cup A_3^{d_1}$, i.e., it can be expressed as
$$
f=\sum_{a_1+a_2=d_1,0<a_1,0\leq a_2\leq q-1}\lambda_{a_1,a_2} x_1^{a_1}x_2^{a_2}+\lambda_{0,d_1}x_2^{d_1}, \lambda_{a_1,a_2}\in \fq.
$$
We assume that $\lambda_{0,d_1}\neq 0$. As in the previous result, we must have an expression
\begin{equation}\label{fa3}
f\equiv \sum_{x_0^{a_0}x_1^{a_1}x_2^{a_2}\in A^{d_2}} \gamma_{a_0,a_1,a_2}x_0^{a_0}x_1^{a_1}x_2^{a_2}  \bmod I(P^2), \gamma_{a_0,a_1,a_2}\in \fq.
\end{equation}
The only monomials in $A^{d_2}$ with $x_2^{\overline{d_1}}$ in their expression in terms of the basis from Lemma \ref{basePm} are $x_0^{a_0}x_1^{a_1}x_2^{\overline{d_1}}$, for some $0<a_0$, $0<a_1\leq q-1$, such that $a_0+a_1+\overline{d_1}=d_2$, and $x_2^{d_2}$ if $d_1\equiv d_2 \bmod q-1$, which is the case that we do not cover. Therefore, we focus on the first type of monomials, which by Lemma \ref{divisionP2} can be expressed as
\begin{equation}\label{exp22}
x_0^{a_0}x_1^{a_1}x_2^{\overline{d_1}} \equiv x_1^{a_1}x_2^{\overline{d_1}}+(x_0-1)(x_2^{\overline{d_1}}+x_1-1)\bmod I(P^2).
\end{equation}
If $d_1\leq q-1$, we have $\overline{d_1}=d_1$ and the monomials $x_1^{a_1}x_2^{d_1}$ have degree greater than $d_1$. Thus, they cannot be in the support of $f$ and they have to be cancelled in the expression from (\ref{fa3}) if we consider some monomial $x_0^{a_0}x_1^{a_1}x_2^{d_1}$. The only other monomial in $A^{d_2}$ with $x_1^{a_1}x_2^{d_1}$ in its expression from the basis from Lemma \ref{basePm} is $x_1^{a_1+q-1}x_2^{d_1}$ if $d_2=a_1+d_1+q-1$. We have $a_1>0$, which implies $d_2-d_1-(q-1)=a_1>0$. We also have that $a_0+a_1+d_1=d_2$, which means that $a_0=q-1$, and the only monomial that we can consider then is $x_0^{q-1}x_1^{d_2-d_1-(q-1)}x_2^{d_1}$ if $d_2-d_1-(q-1)>0$. From (\ref{exp22}) we obtain
$$
x_2^{d_1}\equiv x_1^{d_2-d_1}x_2^{d_1}-x_0^{q-1}x_1^{d_2-d_1-(q-1)}x_2^{d_1}+(x_0-1)(x_1-1)+x_0x_2^{d_1} \bmod I(P^2).
$$
We have seen that $x_0^{q-1}x_1^{d_2-d_1-(q-1)}x_2^{d_1}$ is the only monomial that we can consider to obtain the monomial $x_2^{d_1}$ in the right hand side of (\ref{fa3}), and we need to consider $x_1^{d_2-d_1}x_2^{d_1}$ with the opposite coefficient to cancel the monomial $x_1^{d_2-d_1-(q-1)}x_2^{d_1}$ (the expression of $x_1^{d_2-d_1}x_2^{d_1}$ in terms of the basis from Lemma \ref{basePm}), i.e., we must have $-\gamma_{q-1,d_2-d_1-(q-1),d_1}=\gamma_{0,d_2-d_1,d_1}=\lambda_{0,d_1}$. We can define in this case $(A^{d_2})^{(1)}=A^{d_2}\setminus \{ x_0^{q-1}x_1^{d_2-d_1-(q-1)}x_2^{d_1},x_1^{d_2-d_1}x_2^{d_1}\}$, and consider
\begin{equation}\label{exp11}
\begin{aligned}
 f^{(1)}=&f-\lambda_{0,d_1}x_2^{d_1}\\
 \equiv& \sum_{x_0^{\alpha_0}x_1^{\alpha_1}x_2^{\alpha_2}\in (A^{d_2})^{(1)}} \gamma_{\alpha_0,\alpha_1,\alpha_2}x_0^{\alpha_0}x_1^{\alpha_1}x_2^{\alpha_2} -\lambda_{0,d_1}\left((x_0-1)(x_1-1)+x_0x_2^{d_1} \right) \bmod I(P^2).
\end{aligned}
\end{equation}
Now $f^{(1)}$ only has monomials from $A_2^{d_1}$ in its support, and we can argue as we did in Lemma \ref{lemaa2} to obtain $f^{(l)}=0$ after $l\geq 0$ steps. Taking into account that $d_1\leq q-1$, we see that the monomials left in the support of $f^{(1)}$ are of the type $x_1^{a_1}x_2^{a_2}$ with $a_1+a_2=d_1$, $0<a_1$, which implies that $a_2\leq d_1-1$. Therefore, in the process of obtaining $f^{(l)}$ we do not need to use the monomials that we have used to obtain $f^{(1)}$ because those monomials have $d_1$ as the exponent for $x_2$. Hence, after $l$ steps we obtain an expression similar to (\ref{exp2}), but with the extra term $-\lambda_{0,d_1}\left((x_0-1)(x_1-1)+x_0x_2^{d_1} \right)$ in the right hand side. The same argument proves that we can only have $\lambda_{d_1-a_2,a_2}\neq 0$ if $a_2\in Y= \{0,1,\dots,\min \{ d_1-1,d_2-q\}\}$. If we are in that situation, then we can cancel all the terms in the second sum of the right hand side in (\ref{exp2}). Thus, in that case, we would obtain a sum of monomials in $(A^{d_2})^{(l)}$ equal to $\lambda_{0,d_1}\left((x_0-1)(x_1-1)+x_0x_2^{d_1} \right)$. This implies that we have the evaluation of $(x_0-1)(x_1-1)+x_0x_2^{d_1}$ in $\PRM_{d_2}(2)$. This is a contradiction because we have the evaluation of $x_0^{d_2-d_1}x_2^{d_1}$ in $\PRM_{d_2}(2)$, and $(x_0-1)(x_1-1)+x_0x_2^{d_1}-x_0^{d_2-d_1}x_2^{d_1}\equiv (x_0-1)(x_1-1)\bmod I(P^2)$, whose evaluation has Hamming weight 1. This means that we cannot have $d_1\leq q-1$ and $x_2^{d_1}$ in the support of $f$ simultaneously. 

On the other hand, if $d_1\geq q$, we consider the following polynomials: 
\begin{equation}\label{polQ}
\begin{aligned}
Q_{d_1,d_2}:=x_2^{d_1}+x_1^{d_1-\overline{d_2}}x_2^{\overline{d_2}}+x_0^{d_1-\overline{d_2}}x_2^{\overline{d_2}}+x_0^{d_1-\overline{d_2}}x_1^{\overline{d_2}-\overline{d_1}}x_2^{\overline{d_1}} \in S_{d_1},\\
Q'_{d_2,d_1}:
=x_2^{d_2}+x_1^{d_2-\overline{d_1}}x_2^{\overline{d_1}}+x_0^{d_2-\overline{d_1}}x_2^{\overline{d_1}}+x_0^{d_2-d_1}x_1^{d_1-\overline{d_2}}x_2^{\overline{d_2}} \in S_{d_2}.
\end{aligned}
\end{equation}
These polynomials are obtained by realising that if $x_2^{d_1}$ is in the support of $f$, then we must also have $x_2^{d_2}$ in (\ref{fa3}) to obtain $f(0,0,1)\neq 0$, and adding monomials to obtain polynomials with the same evaluation at $P^2$, we arrive at the polynomials $Q_{d_1,d_2}$ and $Q'_{d_2,d_1}$. As they have the same evaluation at $P^2$, these polynomials are in the same class in $S/I(P^2)$. This also implies that this class is in $\sd\cap \sdd$ and $Q_{d_1,d_2}$ satisfies the conditions in the statement.
\end{proof}

\begin{ex}\label{running3}
We continue with Example \ref{running2}. We had $q=4\leq d_1<d_2=5$. Thus, by Lemma \ref{lemaa3}, we have the polynomials $Q_{4,5}$ and $Q'_{5,4}$ from (\ref{polQ}) in $S_4/I(P^2)\cap S_5/I(P^2)$:
$$
\begin{aligned}
Q_{4,5}&=x_2^4+x_1^2x_2^2+x_0^2x_2^2+x_0^2x_1x_2,\\
Q'_{5,4}&=x_2^5+x_1^4x_2+x_0^4x_2+x_0x_1^2x_2^2.
\end{aligned}
$$
It is easy to check that both polynomials have the same evaluation at $[\{1\}\times \F_4^2]$ as $x_2+x_2^2+x_1x_2+x_1^2x_2^2$, the same evaluation at $[\{0\}\times \{1\}\times \F_4]$ as $x_2+x_2^2$, and both evaluate to 1 at $[0:0:1]$. Therefore, they have the same evaluation at $P^2$.
\end{ex}

With the notation as above, we present the main result of this section.

\begin{thm}\label{basehulleu}
Let $1\leq d_1<d_2\leq 2(q-1)$, and let $Y=\{0,1,\dots,\min\{d_1-1,d_2-q\}\}$. If $d_1\equiv d_2 \bmod q-1$, $A^{d_1}$ is a basis for $\sd\cap \sdd$. If $d_1\not\equiv d_2 \bmod q-1$, the following set $B$ is a basis for $\sd\cap \sdd$:
$$
B=
\begin{cases}
A_1^{d_1} &\text{ if } d_2\leq q-1, \\
A_1^{d_1}\cup \left(\bigcup_{a_2\in Y}\{ x_1^{d_1-a_2}x_2^{a_2} \}\right) &\text{ if } d_1\leq q-1 <d_2, \\
A_1^{d_1}\cup \left(\bigcup_{a_2\in Y}\{ x_1^{d_1-a_2}x_2^{a_2} \}\right)\cup \{ Q_{d_1,d_2}\} &\text{ if } q\leq d_1,
\end{cases}
$$
with $Q_{d_1,d_2}$ defined as in (\ref{polQ}). In particular, the image by the evaluation map of $B$ is a basis for $\PRM_{d_1}(2)\cap \PRM_{d_2}(2)$.
\end{thm}
\begin{proof}
The case $d_1\equiv d_2\bmod q-1$ is covered by Lemma \ref{hullfacilprm}. We assume $d_1\not\equiv d_2\bmod q-1$ now. First, we are going to see that the set $A_1^{d_1}\cup \left(\bigcup_{a_2\in Y}\{ x_1^{d_1-a_2}x_2^{a_2} \}\right)\cup \{ Q_{d_1,d_2}\}$ is linearly independent in $S/I(P^2)$, which proves that all the sets we are considering are linearly independent. $A_1^{d_1}\cup \left(\bigcup_{a_2\in Y}\{ x_1^{d_1-a_2}x_2^{a_2} \}\right)$ is linearly independent because it is a subset of $A^{d_1}$, which is linearly independent. And, when we consider the union with $\{ Q_{d_1,d_2}\}$, we preserve linear independence because $Q_{d_1,d_2}$ is the only polynomial of this union that has nonzero evaluation in $[0:0:1]$, which implies that its evaluation is linearly independent from the rest. On the other hand, these sets are clearly contained in $\sd$, and the proofs from Lemmas \ref{lemaA1}, \ref{lemaa2} and \ref{lemaa3} show that these sets are also contained in $\sdd$.

Hence, we only need to prove that $B$ is a system of generators for $\sd \cap \sdd$. Let $f\in \sd\cap \sdd$. If $f(0,0,1)=\lambda \neq 0$, then $f$ has $x_2^{d_1}$ in its support when expressed in terms of the monomials in $A^{d_1}$. By Lemma \ref{lemaa3}, we must have $q\leq d_1$. Moreover, we can subtract $\lambda Q_{d_1,d_2}$ from $f$ and obtain a polynomial $f^{(1)}$ such that $f^{(1)}(0,0,1)=0$, its expression in terms of the monomials in $A^{d_1}$ only contains monomials from $A_1^{d_1}\cup A_2^{d_1}$, and $f^{(1)}\in \sd \cap \sdd$. On the other hand, if $f^{(1)}$ has monomials from $A_1^{d_1}$ in its support when expressed in terms of the monomials in $A^{d_1}$, by Lemma \ref{lemaA1} we know that we can subtract adequate multiples of those monomials and obtain a polynomial $f^{(2)}\in \sd \cap \sdd$ that only has monomials from $A_2^{d_1}$ in its support. Finally, we can apply Lemma \ref{lemaa2} to $f^{(2)}$ and obtain that $f^{(2)}$ can be generated by $\bigcup_{a_2\in Y}\{ x_1^{d_1-a_2}x_2^{a_2} \}$. If $f(0,0,1)=0$, we can apply the reasoning we have used above for $f^{(1)}$.
\end{proof}

Note that the basis from the previous result is formed by monomials, except when $q\leq d_1$, where we consider $Q_{d_1,d_2}$. This polynomial cannot be reduced to a monomial subtracting other monomials from the basis $B$ of Theorem \ref{basehulleu} because both $x_2^{d_1}$ and $x_1^{d_1-\overline{d_2}}x_2^{\overline{d_2}}$ are linearly independent from the rest of monomials from $B$ by Lemma \ref{basehomogenea} (also check the definition of $Y$ from Lemma \ref{lemaa2} and note that $d_2-q<\overline{d_2}$).

\begin{ex}\label{running4}
Continuing with Examples \ref{running1}, \ref{running2} and \ref{running3}, we see that the set 
$$
A_1^4\cup \{x_1^4,x_1^3x_2\}\cup \{x_2^4+x_1^2x_2^2+x_0^2x_2^2+x_0^2x_1x_2\}
$$
is a basis for $S_4/I(P^2)\cap S_5/I(P^2)$. 
\end{ex}

By counting the elements of the set $B$ in Theorem \ref{basehulleu}, we obtain the dimension of $\sd\cap \sdd$.

\begin{cor}\label{cordim}
Let $1\leq d_1<d_2\leq 2(q-1)$. Let $k_1=\dim\RM_{d_1-1}(2)$. If $d_1\equiv d_2 \bmod q-1$, then $\dim(\PRM_{d_1}(2)\cap \PRM_{d_2}(2))=\dim \PRM_{d_1}(2)$. If $d_1 \not \equiv d_2 \bmod q-1$, then
$$
\dim(\PRM_{d_1}(2)\cap \PRM_{d_2}(2))= 
\begin{cases}
k_1 &\text{ if } d_2\leq q-1,\\
k_1+\min\{d_1,d_2-(q-1)\} &\text{ if } d_1\leq q-1 <d_2, \\
k_1+ d_2-q+2 &\text{ if } q\leq d_1.
\end{cases}
$$
\end{cor}

In the case where $d_2=d_1^\perp=2(q-1)-d_1$, Corollary \ref{cordim} simplies to the following. 

\begin{cor}
Let $1\leq d\leq q-1$, let $Y=\{0,1,\dots,\min\{d-1,q-d-2\}\}$, and let $k_1=\dim \RM_{d-1}(2)$. If $2d\equiv 0\bmod q-1$, then $\PRM_d(2)\cap \PRM_d^\perp(2)=\PRM_d(2)$. If $2d\not\equiv 0\bmod q-1$, a basis for $\PRM_d(2)\cap \PRM_d^\perp(2)$ is given by $A_1^{d}\cup \left(\bigcup_{a_2\in Y}\{ x_1^{d-a_2}x_2^{a_2} \}\right)$. Consequently, $\dim \PRM_d(2)\cap \PRM_d^\perp(2)=k_1+\min\{d,q-d-1\}$.
\end{cor}
\begin{proof}
For the case $d=q-1$, we have that $\PRM_{q-1}(2)=\PRM_{q-1}(2)\cap \PRM_{q-1}^\perp(2)$ by Theorem \ref{dualPRM}. For $1\leq d<q-1$, from Theorem \ref{dualPRM} we see that $\PRM_d^\perp(2)=\PRM_{d^\perp}(2)$, with $d^\perp=2(q-1)-d$. The result is obtained by applying the previous results with $d_1=d$, $d_2=d^\perp$. 
\end{proof}

Note that for $q\leq d< 2(q-1)$, we can also obtain the dimension of the hull by considering $\PRM_{d^\perp}(2)$ in the previous result. For $d=2(q-1)$, by Theorem 
\ref{dualPRM} the dual code is generated by the evaluation of 1, and by Lemma \ref{noesta1} we obtain $\PRM_{2(q-1)}(2)\cap \PRM_{2(q-1)}^\perp(2)=\{0\}$. 

\subsection{Hermitian hull}\label{sechullhermitico}
In the Hermitian case, we consider codes defined over $\F_{q^2}^n$, and the Hermitian product of two vectors $v,w\in \F_{q^2}^n$ is 
$$
v \cdot_h w=\sum_{i=1}^n v_iw_i^q.
$$
The Hermitian dual of a code $C\subset \F_{q^2}^n$ is defined as $C^{\perp_h}:=\{v\in \F_{q^2}^n\mid v\cdot_h w=0, \;\forall\; w \in C \}$. We recall that we defined the Hermitian hull as $\text{Hull}^H(C)=C\cap C^{\perp_h}$. It is easy to check that, for a code $C\subset \F_{q^2}^n$, we have that $C^{\perp_h}=(C^{\perp})^q$, where we consider the component wise power of $q$. In particular, this implies that the Hermitian dual and the Euclidean dual have the same parameters. In this section we show that we may apply similar techniques to the ones used in the previous section to compute the Hermitian hull in some cases. In what follows, as we are working over $\F_{q^2}$, we change $q$ by $q^2$ in the definitions of $S$, $A_i^d$, for $i=1,2,3$, etc. We show now how the main definitions from the other sections are adapted to the Hermitian case in this section:
\begin{enumerate}
    \item $S=\F_{q^2}[x_0,x_1,x_2]$.
    \item Projective and affine Reed-Muller codes are defined for $1\leq d \leq m(q^2-1)$.
    \item $A^d$ is defined for $1\leq d \leq 2(q^2-1)$. If we have $x_0^{a_0}x_1^{a_1}x_2^{a_2}\in A_1^d$, then $0\leq a_1,a_2\leq q^2-1$, and if $x_0^{a_0}x_1^{a_1}x_2^{a_2}\in A_2^d$, then $0\leq a_2\leq q^2-1$. For Lemma \ref{basePm}, we have $0\leq a_i\leq q^2-1$, for $1\leq i \leq 2$.
    \item Now $\overline{z}$ is the integer $1\leq \overline{z}\leq q^2-1$ such that $\overline{z}\equiv z\bmod q^2-1$ when $z>0$, and $\overline{z}=0$ otherwise. 
\end{enumerate}

In the affine case, Reed-Muller codes are either contained in their Euclidean dual or they contain it, which means that the computation of the Euclidean hull is trivial. However, the following result from \cite{galindostabilizer} remarks that computing the Hermitian hull is more difficult than computing the Euclidean hull in the affine case.

\begin{prop}\label{afinhermitico}
The codes' inclusion $\RM_d(q^2,m)\subset \RM_d^{\perp_h}(q^2,m)$ holds if, and only if, $0\leq d\leq m(q-1)-1$.
\end{prop}

Moreover, it is not hard to obtain a basis for the intersection of a Reed-Muller code with the Hermitian dual of another Reed-Muller code. 

\begin{defn}\label{D:Ud1d2}
Let $0\leq d_1,d_2\leq m(q^2-1)$. We define
$$
U_{d_1,d_2}:=\{x_1^{a_1}x_2^{a_2}\mid 0\leq a_1,a_2\leq q^2-1,\; a_1+a_2\leq d_1,\; \overline{qa_1}+\overline{qa_2}\leq 2(q^2-1)-d_2-1\}.
$$
\end{defn}

\begin{prop}\label{hullafin}
The image by the evaluation map over $\mathbb{A}^2$ of $U_{d_1,d_2}$ is a basis for $\RM_{d_1}(q^2,2)\cap\RM_{d_2}^{\perp_h}(q^2,2)$.
\end{prop}
\begin{proof}
The monomials $x_1^{a_1}x_2^{a_2}$ with $0\leq a_1,a_2\leq q^2-1$ have linearly independent evaluations over $\mathbb{A}^2$, and their evaluations generate $\F_{q^2}^{q^4}$ (the full code). The evaluation of a monomial $x_1^{a_1}x_2^{a_2}$ with $0\leq a_1,a_2\leq q^2-1$ and $a_1+a_2\leq d_1$ is in $\RM_{d_2}^{\perp_h}(q^2,2)$ if and only if $x_1^{a_1}x_2^{a_2}\equiv (x_1^{b_1}x_2^{b_2})^q \bmod I(\mathbb{A}^2)$ for some $0\leq b_1,b_2\leq q^2-1$ such that $b_1+b_2\leq 2(q^2-1)-d_2-1$, where $I(\mathbb{A}^2)=\langle x_1^{q^2}-x_1,x_2^{q^2}-x_2\rangle$ (we have used the duality from Theorem \ref{dualRM} and the fact that $\RM_{d_2}^{\perp_h}(q^2,2)=(\RM_{d_2}^{\perp}(q^2,2))^q$). If $a_i\neq 0$ for some $i=1,2$, then $x_1^{a_1}x_2^{a_2}\equiv (x_1^{b_1}x_2^{b_2})^q \bmod I(\mathbb{A}^2)$ implies $a_i\equiv q b_i \bmod q^2-1$ with $b_i\neq 0$, which is equivalent to having $b_i=\overline{qa_i}$ (recall Remark \ref{R:x0}). If $a_i=0$ for some $i=1,2$, then $x_1^{a_1}x_2^{a_2}\equiv (x_1^{b_1}x_2^{b_2})^q \bmod I(\mathbb{A}^2)$ implies $b_i=0=\overline{qa_i}$ in this case as well. Therefore, in both cases $b_i=\overline{qa_i}$, which finishes the proof.
\end{proof}

\begin{rem}
The previous result can be extended in the obvious way to the Reed-Muller codes in $m$ variables. For the Hermitian hull of affine Reed-Muller codes we only need to consider $U_{d,d}$, but for projective Reed-Muller codes we will also consider $U_{d-1,d}$, and that is why we expressed Proposition \ref{hullafin} in full generality with two degrees $d_1$ and $d_2$. 
\end{rem}

Using that $C^{\perp_h}=(C^{\perp})^q$, if we consider $(A_i^d)^q:=\{(x^\alpha)^q \mid x^\alpha \in A_i^d\}$ and $d^\perp=2(q^2-1)-d$, we have that the image by the evaluation map of $(A^{d^\perp})^q:=\bigcup_{i=1}^3(A_i^{d^\perp})^q$ is a basis for $\PRM_d^{\perp_h}(q^2,2)=(\PRM_{d^\perp}(q^2,2))^q$ (if $d\not\equiv 0 \bmod q^2-1$). Following the notation from the previous section, we will denote by $S_{d^\perp}^q/I(P^2)$ the vector space generated in $S/I(P^2)$ by $\bigcup_{i=1}^3(A_i^{d^\perp})^q$.

To compute the dimension of the Hermitian hull of projective Reed-Muller codes it is enough to consider the case with $d\leq q^2-1$. This is because if we assume $d>q^2-1$, then
$$
\PRM_d(q^2,2)\cap \PRM_{d^\perp}^q(q^2,2)=(\PRM_d^q(q^2,2)\cap \PRM_{d^\perp}(q^2,2))^q,
$$
and then at the right-hand side of the previous equality we have the Hermitian hull of a projective Reed-Muller code of degree $d^\perp <q^2-1$, to the power of $q$. Moreover, because of Theorem \ref{dualPRM} we are going to avoid the case with $d=q^2-1$ when giving results for the Hermitian hull (for results about $\sdh\cap \sddh$ we will still consider $d=q^2-1$). This is because in that case $\PRM_d^\perp(q^2,2)=\PRM_{d^\perp}(q^2,2)+\langle (1,\dots,1)\rangle\neq \PRM_{d^\perp}(q^2,2)$, and $\PRM_d^{\perp_h}(q^2,2)=(\PRM_{d^\perp}(q^2,2)+\langle (1,\dots,1)\rangle)^q\neq \PRM_{d^\perp}^q(q^2,2)\cong S_{d^\perp}^q/I(P^2)$. Hence, if $d=q^2-1$ we do not have the isomorphism between $\PRM_d^{\perp_h}(q^2,2)$ and $S_{d^\perp}^q/I(P^2)$, and also this case is the least interesting for quantum codes because we do not have a bound for the minimum distance of the dual code. 

As a consequence of Proposition \ref{afinhermitico} and Proposition \ref{hullafin}, we have the following result. 

\begin{lem}\label{A1hermitico}
Let $1\leq d \leq q^2-1$ and let $U:=\{x_0^{d-a_1-a_2}x_1^{a_1}x_2^{a_2} \mid x_1^{a_1}x_2^{a_2} \in U_{d-1,d}\}\subset S_d$. Then, the classes of the monomials in $U$ are contained in $S_{d^\perp}^q/I(P^2)$. Moreover, if $d\leq 2(q-1)$, $U=A_1^d$.
\end{lem}
\begin{proof}
Let $1\leq d \leq q^2-1$, and let $x_0^{a_0}x_1^{a_1}x_2^{a_2}\in U$. By definition, it is clear that $x_0^{a_0}x_1^{a_1}x_2^{a_2}\in S_{d^\perp}^q/I(P^2)$ because $x_0^{a_0}x_1^{a_1}x_2^{a_2} \equiv (x_0^{d^\perp-\overline{qa_1}-\overline{qa_2}}x_1^{\overline{qa_1}}x_2^{\overline{qa_2}})^q \bmod I(P^2)$, where $\overline{qa_1}+\overline{qa_2}\leq d^\perp-1$ by the definition of $U_{d-1,d}$. If $d\leq 2(q-1)$, we consider $x_0^{a_0}x_1^{a_2}x_2^{a_2}\in A_1^d$. Therefore, $a_1+a_2\leq d-1$ and we have that
$$
\overline{qa_1}+\overline{qa_2}\leq qa_1+qa_2\leq q(d-1)\leq 2(q^2-1)-2(q-1)-q\leq 2(q^2-1)-d-1.
$$
This means that $A_1^d\subset U$ in this case, and the other contention always holds.
\end{proof}

\begin{ex}\label{runningh1}
We consider $q=3$ and $d=7$. Hence, we work over $\F_{3^2}$ and $d>2(q-1)=4$ in this case. One can check that we have
$$
U=A_1^7\setminus \{ x_{0} x_{1} x_{2}^{5}, x_{0} x_{1}^{2} x_{2}^{4}, x_{0} x_{1}^{4} x_{2}^{2}, x_{0} x_{1}^{5} x_{2}, x_{0}^{3} x_{1}^{2} x_{2}^{2}, x_{0}^{4} x_{1} x_{2}^{2}, x_{0}^{4} x_{1}^{2} x_{2}\},
$$
where $A_1^7$ is formed by all the monomials of degree $7$ that are divisible by $x_0$ in this case. For instance, for the monomial $x_0x_1x_2^5$ we have $a_1=1$, $a_2=5$, and we check
$$
\overline{qa_1}+\overline{qa_2}=\overline{3}+\overline{15}=10\not \leq 9=d^\perp,
$$
which implies $x_0x_1x_2^5\not \in U$.
\end{ex}

\begin{rem}\label{remhullafin}
To compute the dimension of the hull of projective Reed-Muller codes we will need the size of the set $U$ from Lemma \ref{A1hermitico}. We give a combinatorial formula for $\abs{U}$ in Lemma \ref{lemacontarU} of the Appendix. Moreover, it is possible to obtain a combinatorial formula for $\abs{U_{d,d}}$ as well, as we note in Remark \ref{remdimhullhermafin}, which gives the dimension of the Hermitian hull for affine Reed-Muller codes in 2 variables. 
\end{rem}

In the following results we argue in a similar way to Section \ref{seceu} to show which polynomials can be in $\sdh\cap \sddh$ depending on whether the monomials in the support of these polynomials are contained in $A_2^d$ or if they have $x_2^d$ in their support (we recall that $A_3^d=\{x_2^d\}$). We restrict to the case $1\leq d\leq 2(q-1)$ in some results because in that case we have $U=A_1^d$ by Lemma \ref{A1hermitico}, which is similar to what happens in the Euclidean case. For the following results, recall that $\overline{z}$ is a representative of the class of $z$ modulo $q^2-1$.

\begin{lem}\label{A2hermitico}
Let $1\leq d \leq 2(q-1)$. Let $f\in S_{d}$ such that it only has monomials from $ A_2^{d}$ in its support, i.e., it can be expressed as
$$
f=\sum_{a_1+a_2=d,0<a_1,0\leq a_2\leq q^2-1}\lambda_{a_1,a_2} x_1^{a_1}x_2^{a_2}, \lambda_{a_1,a_2}\in \F_{q^2}.
$$
Let $T=\{a_2\mid a_2<d,\; d^\perp>\overline{qa_2}+(q^2-1)\}$ and $T_f=\{0\leq a_2\leq q-1 \mid \lambda_{d-a_2,a_2}\neq 0\}$. Then $f\in \sddh$ if and only if $d\equiv 0 \bmod q-1$ or $T_f\subset T$. 
\end{lem}
\begin{proof}
Assuming that $f\in \sddh$, there is an expression
\begin{equation}\label{fa2h}
f\equiv \sum_{(x_0^{\alpha_0}x_1^{\alpha_1}x_2^{\alpha_2})^q\in (A^{d^\perp})^q} \mu_{\alpha_0,\alpha_1,\alpha_2}(x_0^{\alpha_0}x_1^{\alpha_1}x_2^{\alpha_2})^q  \bmod I(P^2), \mu_{\alpha_0,\alpha_1,\alpha_2}\in \F_{q^2}.
\end{equation}
We have $f(0,0,1)=0$, which means that $\mu_{0,0,d^\perp}=0$. Following the proof of Lemma \ref{lemaa2}, we consider $\ini(f)=x_1^{a_1}x_2^{a_2}$, with $a_1>0$, $0\leq a_2\leq q^2-1$ and $a_1+a_2=d$ (since $d\leq 2(q-1)$, we also have $a_1\leq q^2-1$). Because of (\ref{fa2h}) we must have some monomial in $(A^{d^\perp})^q$ such that its expression in the basis from Lemma \ref{basePm} contains $x_1^{a_1}x_2^{a_2}$ in its support. Let $\gamma_i=\overline{qa_i}$, for $i=1,2$, which implies that $q\gamma_i\equiv a_i\bmod q^2-1$. The only monomials in $(A^{d^\perp})^q$ that contain $x_1^{a_1}x_2^{a_2}$ in their expression in terms of the basis from Lemma \ref{basePm} are $(x_1^{d^\perp-\gamma_2}x_2^{\gamma_2})^q$ if $q(d^\perp-\gamma_2)\equiv a_1\bmod q^2-1$, and $(x_0^{\gamma_0}x_1^{\gamma_1}x_2^{\gamma_2})^q$ if $\gamma_0=d^\perp-\gamma_1-\gamma_2>0$.

In the first case, $q(d^\perp-\gamma_2)\equiv a_1\bmod q^2-1$ implies that $qd^\perp\equiv d \bmod q^2-1$, which happens if and only if $d\equiv 0 \bmod q-1$. Taking into account that $d\leq 2(q-1)$, we have $d^\perp>q^2-1\geq \gamma_2$, and we have $x_1^{a_1}x_2^{a_2}\equiv (x_1^{d^\perp-\gamma_2}x_2^{\gamma_2})^q \bmod I(P^2)$. Moreover, in this situation we can do this for all the monomials from $A_2^d$ in the support of $f$.

If $d\not \equiv 0 \bmod q-1$, the only monomial in $(A^{d^\perp})^q$ with $x_1^{a_1}x_2^{a_2}$ in its expression in terms of the elements of the basis from Lemma \ref{basePm} is $(x_0^{\gamma_0}x_1^{\gamma_1}x_2^{\gamma_2})^q$ with $\gamma_0=d^\perp -\gamma_1-\gamma_2$, if $d^\perp -\gamma_1-\gamma_2>0$. This is because, by Lemma \ref{divisionP2}, we have that
$$
(x_0^{\gamma_0}x_1^{\gamma_1}x_2^{\gamma_2})^q\equiv x_1^{a_1}x_2^{a_2}+(x_0-1)(x_2^{a_2}+x_1-1)\bmod I(P^2),
$$
if $\gamma_0>0$. Hence, we must have $\mu_{\gamma_0,\gamma_1,\gamma_2}=\lambda_{a_1,a_2}$. If we denote by $(\adq)^{(1)}=\adq \setminus \{ (x_0^{\gamma_0}x_1^{\gamma_1}x_2^{\gamma_2})^q\}$, we obtain 

\begin{equation}\label{exp1h}
\begin{aligned}
 f^{(1)}=&f-\lambda_{a_1,a_2}x_1^{a_1}x_2^{a_2}\\
 \equiv& \sum_{(x_0^{\alpha_0}x_1^{\alpha_1}x_2^{\alpha_2})^q\in (\adq)^{(1)}} \mu_{\alpha_0,\alpha_1,\alpha_2}(x_0^{\alpha_0}x_1^{\alpha_1}x_2^{\alpha_2})^q\\
 &+\mu_{d^\perp-\overline{qa_1}-\overline{qa_2},\overline{qa_1},\overline{qa_2}}(x_0-1)(x_2^{a_2}+x_1-1) \bmod I(P^2).
\end{aligned}
\end{equation}

Arguing as in the proof of Lemma \ref{lemaa2}, after $l$ steps we get
\begin{equation}\label{exp2h}
\begin{aligned}
0\equiv& \sum_{(x_0^{\alpha_0}x_1^{\alpha_1}x_2^{\alpha_2})^q\in (\adq)^{(l)}} \mu_{\alpha_0,\alpha_1,\alpha_2}(x_0^{\alpha_0}x_1^{\alpha_1}x_2^{\alpha_2})^q \\
&+(x_0-1)\sum_{a_2\in T_f}\mu_{d^\perp-\overline{q(d-a_2)}-\overline{qa_2},\overline{q(d-a_2)},\overline{qa_2}}(x_2^{a_2}+x_1-1) \bmod I(P^2),
\end{aligned}
\end{equation}
where $T_f=\{0\leq a_2\leq q^2-1 \mid \lambda_{d-a_2,a_2}\neq 0\}$. With this notation, we have that $(\adq)^{(l)}=\adq\setminus \bigcup_{a_2\in T}\{ (x_0^{\gamma_0}x_1^{\gamma_1}x_2^{\gamma_2})^q,\gamma_0=d^\perp-\gamma_1-\gamma_2,\;\gamma_1=\overline{q(d-a_2)} ,\;\gamma_2 =\overline{qa_2} \}$. 

If we express all the monomials in (\ref{exp2h}) in terms of the basis from Lemma \ref{basePm}, then we must have the coefficient of each element of the basis equal to 0. The monomials from (\ref{exp2h}) in the second sum are already expressed in terms of the basis from Lemma \ref{basePm}. If we focus on the monomial $x_2^{a_2}$ for some $a_2\in T_f$ with $a_2>0$, we see that all the monomials $(x_0^{c_0}x_1^{c_1}x_2^{\gamma_2})^q\in (\adq)^{(l)}$ with $0<c_0$, $0<c_1\leq q^2-1$, $qc_1\not\equiv d-a_2 \bmod q^2-1$, and $c_0+c_1+\gamma_2=d^\perp$, have $x_2^{a_2}$ in their expression in terms of the basis from Lemma \ref{basePm}:
$$
(x_0^{c_0}x_1^{c_1}x_2^{\gamma_2})^q\equiv x_1^{\overline{qc_1}}x_2^{a_2}+(x_0-1)(x_2^{a_2}+x_1-1) \bmod I(P^2).
$$
In fact, these are the only monomials from $(\adq)^{(l)}$ with $x_2^{a_2}$ in their expression (the one with $q c_1\equiv d-a_2\bmod q^2-1$ is not contained in $(\adq)^{(l)}$). However, if we have $\mu_{c_0,c_1,\gamma_2}\neq 0$, some other monomial from $(\adq)^{(l)}$ must cancel the monomial $x_1^{\overline{qc_1}}x_2^{a_2}$ from (\ref{exp2h}). The only other monomial in $(\adq)^{(l)}$ with $x_1^{\overline{qc_1}}x_2^{a_2}$ in its support when expressed in terms of the basis from Lemma \ref{basePm} is $(x_1^{c_1+q^2-1}x_2^{\gamma_2})^q$ (we cannot use $(x_1^{c_1}x_2^{\gamma_2})^q$ because $c_1+\gamma_2<d^\perp$), if $c_1+\gamma_2+q^2-1=d^\perp$ (which implies $c_0=q^2-1$). In our case, we always have $q^2-1<d^\perp$, but still we must also have $d^\perp>\gamma_2+q^2-1$ to ensure $c_1> 0$. 

Therefore, if $d^\perp-\gamma_2-(q^2-1)=c_1>0$, we can consider the following polynomial in $S_{d^\perp}^q/I(P^2)$:
$$
(x_0^{q^2-1}x_1^{c_1}x_2^{\gamma_2})^q-(x_1^{c_1+q^2-1}x_2^{\gamma_2})^q \equiv (x_0-1)(x_2^{a_2}+x_1-1) \bmod I(P^2).
$$

For every $a_2\in T_f$, we must have $\mu_{q^2-1,c_1,\gamma_2}=-\mu_{0,c_1+q^2-1,\gamma_2}=\lambda_{d-a_2,a_2}$ to cancel the polynomial $(x_0-1)(x_2^{a_2}+x_1-1)$ from (\ref{exp2}). Thus, have seen that $d^\perp>\gamma_2+(q^2-1)$, which implies that $T_f\subset T$.

These are necessary conditions, and now we show that they are sufficient. We assume $T_f\subset T$, and for each $a_2\in T_f$, we denote $\gamma_1=\overline{q(d-a_2)}$ as before. Then we have
\begin{equation}\label{polV}
x_1^{d-a_2}x_2^{a_2}\equiv (x_1^{d^\perp-\gamma_2}x_2^{\gamma_2}-x_0^{q^2-1}x_1^{\overline{d^\perp}-\gamma_2}x_2^{\gamma_2}+x_0^{d^\perp-\gamma_1-\gamma_2}x_1^{\gamma_1}x_2^{\gamma_2})^q \bmod I(P^2).
\end{equation}
We note that if $\overline{d^\perp}=d^\perp-(q^2-1)>\gamma_2$, then $d^\perp>\gamma_1+\gamma_2$. The previous equality is easy to check because both sides have the same evaluation at $P^2=[\{1\}\times \F_{q^2}^2]\cup[\{0\}\times \{1\}\times \F_{q^2}]\cup\{[0:0:1]\}$. Hence, $x_1^{d-a_2}x_2^{a_2}\in \sdh\cap \sddh$ for each $a_2\in T_f$, which implies that $f\in \sdh\cap \sddh$. 
\end{proof}

\begin{lem}\label{A3hermitico}
Let $1\leq d\leq 2(q-1)$. There is some $f\in S_{d}$ with $x_2^{d}$ in its support and such that $f\in \sddh$ if and only if $d\equiv 0 \bmod q-1$.
\end{lem}
\begin{proof}
If $d\equiv 0 \bmod q-1$, then we have $x_2^d\equiv x_2^{qd^\perp}\bmod I(P^2)$ because $$d\equiv qd^\perp\bmod q^2-1\iff (q+1)d\equiv 0 \bmod q^2-1\iff d\equiv 0 \bmod q-1.$$ 
Therefore, we only have to prove the other implication.

Let $f\in S_{d}$. By Lemma \ref{A1hermitico}, we can assume that the support of $f$ is contained in $ (A_2^{d^\perp})^q\cup (A_3^{d^\perp})^q$, i.e., it can be expressed as
$$
f=\sum_{a_1+a_2=d,0<a_1,0\leq a_2\leq q^2-1}\lambda_{a_1,a_2} x_1^{a_1}x_2^{a_2}+\lambda_{0,d}x_2^{d}, \lambda_{a_1,a_2}\in \F_{q^2}.
$$
We assume that $\lambda_{0,d}\neq 0$. As in the previous result, we must have an expression
\begin{equation}\label{fa3h}
f\equiv \sum_{x_0^{a_0}x_1^{a_1}x_2^{a_2}\in (A^{d^\perp})^q} \mu_{a_0,a_1,a_2}x_0^{a_0}x_1^{a_1}x_2^{a_2}  \bmod I(P^2), \mu_{a_0,a_1,a_2}\in \F_{q^2}.
\end{equation}
The only monomials in $(A^{d^\perp})^q$ with $x_2^{d}$ in their expression in terms of the basis from Lemma \ref{basePm} are $(x_0^{\beta_0}x_1^{\beta_1}x_2^{\beta_2})^q$, for some $0<\beta_0$, $0< \beta_1\leq q^2-1$, $0<\beta_2\leq q^2-1$, such that $\beta_0+\beta_1+\beta_2=d^\perp$ and $q\beta_2\equiv d \bmod q^2-1$; and $x_2^{d}$ if $d\equiv 0 \bmod q^2-1$. We assume now that $d\not\equiv 0 \bmod q^2-1$, and we will arrive at a contradiction. Thus, we focus on the first type of monomials, which by Lemma \ref{divisionP2} can be expressed as
\begin{equation}\label{exp22h}
(x_0^{\beta_0}x_1^{\beta_1}x_2^{\beta_2})^q \equiv x_1^{\overline{q\beta_1}}x_2^{d}+(x_0-1)(x_2^d+x_1-1)\bmod I(P^2).
\end{equation}
The monomials $x_1^{\overline{q\beta_1}}x_2^{d}$ have degree greater than $d$. Thus, they cannot be in the support of $f$ and they have to be cancelled in the expression from (\ref{fa3h}) if we consider some monomial $(x_0^{\beta_0}x_1^{\beta_1}x_2^{\beta_2})^q$. The only other monomial in $(A^{d^\perp})^q$ with $x_1^{\overline{q\beta_1}}x_2^{d}$ in its expression from the basis from Lemma \ref{basePm} is $(x_1^{\beta_1+q^2-1}x_2^{\beta_2})^q$ if $d^\perp=\beta_1+\beta_2+q^2-1$, which implies $\beta_0=q^2-1$ and $\beta_1=d^\perp-\beta_2-(q^2-1)$. Thus, there is only one monomial in $(A^{d^\perp})^q$ that we can consider, and we have 
$$
x_2^{d}\equiv (x_1^{\beta_1+q^2-1}x_2^{\beta_2})^q-(x_0^{\beta_0}x_1^{\beta_1}x_2^{\beta_2})^q+(x_0-1)(x_1-1)+x_0x_2^{d} \bmod I(P^2).
$$
Similarly to the proof of Lemma \ref{lemaa2}, we must have $-\mu_{\beta_0,\beta_1,\beta_2}=\mu_{0,\beta_1+q^2-1,\beta_2}=\lambda_{0,d}$. We can define in this case $((A_2^{d^\perp})^q)^{(1)}=(A_2^{d^\perp})^q\setminus \{ (x_0^{\beta_1}x_1^{\beta_1}x_2^{\beta_2})^q,(x_1^{\beta_1+q^2-1}x_2^{\beta_2})^q\}$, and consider
\begin{equation}\label{exp11h}
\begin{aligned}
 f^{(1)}=&f-\lambda_{0,d}x_2^{d}\\
 \equiv& \sum_{x_0^{a_0}x_1^{a_1}x_2^{a_2}\in ((A^{d^\perp})^q)^{(1)}} \mu_{a_0,a_1,a_2}x_0^{a_0}x_1^{a_1}x_2^{a_2} -\lambda_{0,d}\left((x_0-1)(x_1-1)+x_0x_2^{d} \right) \bmod I(P^2).
\end{aligned}
\end{equation}
Arguing as in Lemma \ref{lemaa2} and using Lemma \ref{A1hermitico}, we obtain that there must be a sum of monomials in $\adql$ equal to $\lambda_{0,d}\left((x_0-1)(x_1-1)+x_0x_2^{d} \right)$. This implies that we have the evaluation of $(x_0-1)(x_1-1)+x_0x_2^{d}$ in $\PRM_{d}^{\perp_h}(q^2,2)$. This is a contradiction because we have the evaluation of $(x_0^{d^\perp-\beta_2}x_2^{\beta_2})^q$ in $\PRM_{d}^{\perp_h}(q^2,2)$, and $(x_0-1)(x_1-1)+x_0x_2^{d}-(x_0^{d^\perp-\beta_2}x_2^{\beta_2})^q\equiv (x_0-1)(x_1-1)\bmod I(P^2)$, whose evaluation has Hamming weight 1.
\end{proof}

Let $1\leq d\leq q^2-1$. In the next result we give a basis of $\sdh\cap \sddh$ for $1\leq d\leq 2(q-1)$ using the previous results, and we give a linearly independent set contained in $\sdh\cap \sddh$ for the case $2(q-1)<d\leq q^2-1$. 

To state the next result, we use the following sets of polynomials. We recall that $U=\{x_0^{d-a_1-a_2}x_1^{a_1}x_2^{a_2} \mid x_1^{a_1}x_2^{a_2} \in U_{d-1,d}\}$, where we consider $U_{d-1,d}$ as in Definition \ref{D:Ud1d2}. We define 
$$
V:=\{x_1^{d-a_2}x_2^{a_2}\mid a_2\in T\},
$$
where $T$ is defined as in Lemma \ref{A2hermitico}. Finally, for the case $2(q-1)<d\leq q^2-1$ we define 
$$
\begin{aligned}
W:=\{x_1^{d-a_2}x_2^{a_2}+x_0^{d-\overline{qd^\perp-a_2}-a_2} x_1^{\overline{qd^\perp-a_2}}x_2^{a_2}\mid & q^2-1\geq d^\perp-\overline{qa_2}>\overline{q(d-a_2)} \\
&\text{ and } d-a_2>\overline{qd^\perp -a_2}\}.
\end{aligned}
$$ 

We are interested in $W$ because of the following result.

\begin{lem}\label{L:W}
We have $W\subset \sdh\cap \sddh$, and $U\cup V\cup W$ is a linearly independent set in $S/I(P^2)$.
\end{lem}
\begin{proof}
If  $d^\perp-\overline{qa_2}>\overline{q(d-a_2)}$ and $d-a_2>\overline{qd^\perp -a_2}$ (these conditions come from the definition of $W$), then 
\begin{equation}\label{eqW}
\begin{aligned}
x_1^{d-a_2}x_2^{a_2}+x_0^{d-\overline{qd^\perp-a_2}-a_2}&x_1^{\overline{qd^\perp-a_2}}x_2^{a_2}\equiv \\
&(x_1^{d^\perp-\overline{qa_2}}x_2^{\overline{qa_2}}+x_0^{d^\perp-\overline{q(d-a_2)}-\overline{qa_2}}x_1^{\overline{q(d-a_2)}}x_2^{\overline{qa_2}})^q\bmod I(P^2).
\end{aligned}
\end{equation}
This equivalence can be checked by considering the evaluation of both polynomials at $P^2$. Therefore, we have $W\subset \sdh\cap \sddh$. The condition $q^2-1\geq  d^\perp-\overline{qa_2}$ ensures $U\cup V\cup W$ is linearly independent. Indeed, both of the monomials in the left hand side of (\ref{eqW}) are monomials from the basis of Lemma \ref{basehomogenea} and are not in $U\cup V$. The monomial $x_1^{d-a_2}x_2^{a_2}$ from (\ref{eqW}) is not in $V$ because $d^\perp\leq \overline{qa_2}-(q^2-1)$ (see the definition of $T$ in Lemma \ref{A2hermitico}). For the monomial $x_0^{d-\overline{qd^\perp-a_2}-a_2}x_1^{\overline{qd^\perp-a_2}}x_2^{a_2}$, it is not in $U$ because we have $x_1^{\overline{qd^\perp-a_2}}x_2^{a_2}\not \in U_{d-1,d}$. To see this, we use the definition of $U_{d-1,d}$ from Definition \ref{D:Ud1d2}. First, $d-a_2>\overline{qd^\perp -a_2}$ means that this monomial satisfies the first condition in the definition of $U_{d-1,d}$. For the second one, we would have to check if 
\begin{equation}\label{cond2W}
\overline{q(qd^\perp-a_2)}+\overline{qa_2}=\overline{d^\perp-\overline{qa_2}}+\overline{qa_2}\leq d^\perp-1.
\end{equation}
If $\overline{d^\perp-\overline{qa_2}}<d^\perp-\overline{qa_2}$, we clearly have (\ref{cond2W}). However, if  $\overline{d^\perp-\overline{qa_2}}=d^\perp-\overline{qa_2}$, we do not have (\ref{cond2W}), and this happens if and only if $q^2-1\geq d^\perp -\overline{qa_2}$, which is the condition we are using in the definition of $W$.
\end{proof}

Now we can state the main result of this section. 

\begin{thm}\label{hullproyectivohermitico}
Let $1\leq d \leq  q^2-1$, and $U=\{x_0^{d-a_1-a_2}x_1^{a_1}x_2^{a_2} \mid x_1^{a_1}x_2^{a_2} \in U_{d-1,d}\}$, where we consider $U_{d-1,d}$ as in Definition \ref{D:Ud1d2}. Let $V$ and $W$ be as in the discussion above. If $d\equiv 0 \bmod q-1$, then $U\cup A_2^d\cup A_3^d$ is a basis for $\sdh\cap \sddh$. For $d\not\equiv 0\bmod q-1$, if $d\leq 2(q-1)$, then $U\cup V=A_1^d\cup V$ is a basis for $\sdh\cap \sddh$. Lastly, if $2(q-1)<d\leq q^2-1$, then $U\cup V\cup  W$ is a linearly independent set contained in $\sdh\cap \sddh$.
\end{thm}

\begin{proof}
If $d\equiv 0 \bmod q-1$ and $d\leq q^2-1$, reasoning as in the proofs of Lemmas \ref{A1hermitico}, \ref{A2hermitico} and \ref{A3hermitico}, we have $U\cup A_2^d\cup A_3^d$ contained in $\sdh\cap \sddh$. Let $f\in S_d$ such that $f\in\sddh$ and whose support is contained in $A_1^d\setminus U$. Then $f$ has an expression in terms of the monomials from $\adq$ in $S/I(P^2)$, and this expression only involves the monomials from $(A_1^{d^\perp})^q$. This is because $x_2^{qd^\perp}$ cannot be in the expression because it is nonzero at $[0:0:1]$ while $f(0,0,1)=0$, and if there were monomials from $(A_2^{d^\perp})^q$, in $[\{0\}\times \{1\}\times \F_{q^2}]$ that expression would have the same evaluation as some polynomial in $x_2$ with degree less than or equal to $q^2-1$, which cannot have $q^2$ zeroes ($f$ is equal to $0$ in those points). We finish this case by using the affine case from Proposition \ref{hullafin}.

If $d\not\equiv 0 \bmod q-1$ and $d\leq 2(q-1)$, we have $U=A_1^d$ contained in $\sdh\cap \sddh$ by Lemma \ref{A1hermitico}. Arguing as in the proof of Theorem \ref{basehulleu} and using Lemma \ref{A2hermitico} and Lemma \ref{A3hermitico}, we obtain that $U\cup V$ is a basis for $\sdh\cap \sddh$.

Finally, in the case $d\not \equiv 0\bmod q-1$ and $2(q-1)<d<q^2-1$, we have from Lemma \ref{A1hermitico} that $U$ is contained in $\sdh\cap \sddh$. The fact that $V$ is contained in $\sdh\cap \sddh$ follows from (\ref{polV}). For $W$, by Lemma \ref{L:W}, we have that $W\subset \sdh\cap \sddh$ and $U\cup V\cup W$ is linearly independent in $S/I(P^2)$.
\end{proof}

From Theorem \ref{hullproyectivohermitico} we can obtain the exact dimension of $\sdh\cap \sddh$ if $d\leq 2(q-1)$ or $d\equiv 0 \bmod q-1$. For $2(q-1)<d\leq q^2-1$, $d\not\equiv 0\bmod q-1$, we only have a lower bound for the dimension of $\sdh\cap \sddh$. Note that a lower bound for the dimension of the hull gives an upper bound for the parameter $c$ of the corresponding EAQECC, which is still interesting because it tells us how many maximally entangled pairs are required at most for using that EAQECC. Equivalently, if we use as many maximally entangled pairs as the bound specifies, then we can employ this EAQECC. Nevertheless, in all cases we have checked, this is indeed the true value of the dimension of the hull, 
which implies that $U\cup V\cup W$ is also a basis for the hull in those cases. In particular, this means that the Hermitian hull is not generated by monomials in general because of $W$ (we saw in the proof of Lemma \ref{L:W} that no monomial of the polynomials from $W$ is contained in $U\cup V$). We see this in the next example.

\begin{ex}\label{runningh2}
We continue with the setting from Example \ref{runningh1}. We have $d=7>2(q-1)$ and $d\not\equiv 0 \bmod q-1$. Therefore, by Theorem \ref{hullproyectivohermitico}, $U\cup V\cup W$ is a linearly independent set contained in $\sdh\cap \sddh$. In Example \ref{runningh1} we computed the set $U$, and we are going to compute the sets $V$ and $W$ now.

For $V$, we first obtain $T=\{7\}$. This is because $d^\perp-(q^2-1)=1$ in this case. Thus, $1>\overline{qa_2}$ implies $a_2=0$, and we have $V=\{x_1^7\}$. 

Finally, for $W$, we have to consider $0\leq a_2\leq 7$ and check the conditions in the definition of $W$. In this case, the only $a_2$ that satisfies the conditions is $a_2=1$, and we have $W=\{ x_1^6x_2+x_0^4x_1^2x_2\}$. 

It can be checked with Magma \cite{magma} that the image by the evaluation map of $U\cup V\cup W$ is, in fact, a basis for the Hermitian hull in this case. We also see that the monomials from the polynomial $x_1^6x_2+x_0^4x_1^2x_2$ in the set $W$ are not contained in $V$ and $U$ (see Example \ref{runningh1}). Hence, we see that in this case the Hermitian hull cannot be generated by monomials from $A^d$. 
\end{ex}

We have the following lemma, which gives us the size of the set $T$ (which is the same as the size of the set $V$ as defined prior to Theorem \ref{hullproyectivohermitico}) and allows us to give more explicit expressions for the dimension of $\sdh\cap \sddh$ in some cases.

\begin{lem}\label{lemT}
Let $1\leq d \leq q^2-1$, and let $d=\beta_0+\beta_1 q$ be its $q$-adic expansion. Then 
$$
\abs{T}=\abs{V}=\beta_1(q-1-\beta_1)+\min\{\beta_0,q-1-\beta_1\}+\min\{\beta_1,q-1-\beta_0\}.
$$
\end{lem}
\begin{proof}
Let $a_2\in T$, and we consider its $q$-adic expansion $a_2=\alpha_0+\alpha_1q$. We must have $a_2<d$ and $d^\perp>\overline{qa_2}+q^2-1$ by the definition of $T$. It is easy to check that $\alpha_1+\alpha_0q$ is the $q$-adic expansion of $\overline{qa_2}$. The condition $a_2<d$ translates to the condition
\begin{equation}\label{cond1}
\alpha_1<\beta_1 \text{ or } \alpha_1=\beta_1 \text{ and } \alpha_0<\beta_0.
\end{equation}
For the other condition, it is easy to check that $q-1-\beta_0+(q-1-\beta_1)q$ is the $q$-adic expansion of $d^\perp-(q^2-1)=q^2-1-d$ (using $q^2=q-1+(q-1)q$). Then, the condition $d^\perp-(q^2-1)>\overline{qa_2}$ translates to 
\begin{equation}\label{cond2}
\alpha_0<q-1-\beta_1 \text{ or } \alpha_0=q-1-\beta_1 \text{ and } \alpha_1<q-1-\beta_0.
\end{equation}
Now we count all the pairs $0\leq \alpha_0,\alpha_1\leq q-1$ that satisfy the conditions (\ref{cond1}) and (\ref{cond2}). If $\alpha_0<q-1-\beta_1$, all the values of $\alpha_1$ such that $\alpha_1<\beta_1$ satisfy the conditions. We obtain $\beta_1(q-1-\beta_1)$ pairs in this way.

If $\alpha_0<q-1-\beta_1$, we also have the possibility of having $\alpha_1=\beta_1$, but then we must also have $\alpha_0<\beta_0$ by (\ref{cond1}). Therefore, we obtain the pairs with $\alpha_1=\beta_1$, and $\alpha_0=0,1,\dots,\min\{\beta_0-1,q-2-\beta_1 \}$, i.e., we obtain $\min \{ \beta_0,q-1-\beta_1\}$ pairs of this type. 

If $\alpha_0=q-1-\beta_1$, we must have $\alpha_1<q-1-\beta_0$ by (\ref{cond2}). If we also have $\alpha_1<\beta_1$, we satisfy (\ref{cond1}) and we obtain $\min \{\beta_1,q-1-\beta_0\}$ pairs. The last option would be to have $\alpha_0=q-1-\beta_1$ and $\alpha_1=\beta_1$, in which case we must also have $\alpha_1=\beta_1<q-1-\beta_0$ by (\ref{cond2}) and $\alpha_0=q-1-\beta_1<\beta_0$ by (\ref{cond1}). But we cannot have $\beta_1<q-1-\beta_0$ and $q-1-\beta_1<\beta_0$ simultaneously, which means that this pair does not satisfy the conditions. 
\end{proof}

\begin{rem}\label{remT}
In the previous result, we have that $\beta_1\leq q-1-\beta_0 \iff \beta_0\leq q-1-\beta_1 \iff \beta_0+\beta_1\leq q-1$. Hence, the size of the set $T$ can also be expressed as 
$$
\abs{T}=
\begin{cases}
    \beta_1(q-1-\beta_1)+\beta_0+\beta_1 &\text{ if } \beta_0+\beta_1\leq q-1,\\
    \beta_1(q-1-\beta_1)+2(q-1)-\beta_0-\beta_1 &\text{ if } \beta_0+\beta_1> q-1.
\end{cases}
$$
Moreover, it is easy to check that these expressions can also be written in the following way:
$$
\abs{T}=
\begin{cases}
    d-\beta_1^2 &\text{ if } \beta_0+\beta_1\leq q-1,\\
    d-\beta_1^2 -2(\beta_0+\beta_1-(q-1)) &\text{ if } \beta_0+\beta_1> q-1.
\end{cases}
$$
We note that $d\leq 2(q-1)=q+(q-2)$ implies that $\beta_0+\beta_1\leq q-1$.
\end{rem}

\begin{ex}\label{runningh3}
Continuing with the setting from Example \ref{runningh2}, we have that $d=7=1+2\cdot 3$, which means that $\beta_0=1$, $\beta_1=2$. Thus, by Lemma \ref{lemT}, we obtain
$$
\abs{T}=d-\beta_1^2-2(\beta_0+\beta_1-(q-1))=1,
$$
which is what we obtained in Example \ref{runningh2}.
\end{ex}

As a consequence of Lemma \ref{A1hermitico}, Theorem \ref{hullproyectivohermitico} and Lemma \ref{lemT} we have the following result about the dimension of the Hermitian hull. Note that $\abs{V}=\abs{T}$, and $\abs{U}$ is computed in Lemma \ref{lemacontarU}.

\begin{cor}\label{dimhullherm}
Let $1\leq d < q^2-1$, and let $d=\beta_0+\beta_1q$ be its $q$-adic expansion. If $d\equiv 0 \bmod q-1$, we have
$$
 \dim (\PRM_d(q^2,2)\cap \PRM_d^{\perp_h}(q^2,2))=
\begin{cases}
    \dim(\PRM_d(q^2,2)) &\text{ if } d\leq 2(q-1), \\
    \abs{U}+d+1 &\text{ if }d>2(q-1).
\end{cases}
$$
For the case $d\not\equiv 0 \bmod q-1$: if $d\leq 2(q-1)$, we have
$$
 \dim (\PRM_d(q^2,2)\cap \PRM_d^{\perp_h}(q^2,2))=\abs{U}+d-\beta_1^2=\abs{A_1^d}+d-\beta_1^2,
$$
and if $d>2(q-1)$, we have the lower bound
$$
 \dim (\PRM_d(q^2,2)\cap \PRM_d^{\perp_h}(q^2,2)) \geq \abs{U}+\abs{V} +\abs{W}.
$$
\end{cor}

\begin{ex}
We continue with the setting from Example \ref{runningh2}, where we saw that $U\cup V\cup W$ was a basis for the Hermitian hull. Therefore, we have that the dimension of the Hermitian hull is $\abs{U}+\abs{V}+\abs{W}$ (the bound from Corollary \ref{dimhullherm} is, in fact, the true dimension). From Example \ref{runningh1} we obtain $\abs{U}=28-7=21$, and from Example \ref{runningh2} we obtain $\abs{V}=\abs{W}=1$. Hence, the dimension of the Hermitian hull in this case is $23$. 
\end{ex}

\section{Quantum codes from projective Reed-Muller codes}\label{secquant}
This section is devoted to providing the parameters of the EAQECCs obtained by using projective Reed-Muller codes over the projective plane $\mathbb{P}^2$. Note that, by Theorem \ref{paramPRM} and Theorem \ref{dualPRM}, we know all the parameters of the projective Reed-Muller codes except when $d\equiv 0 \bmod q-1$ (resp. $d\equiv 0 \bmod q^2-1$ in the Hermitian case), in which case we do not know the minimum distance of the dual code. Moreover, in this case the dimension of the hull is not directly given by the computations made in the previous sections because we would have to also consider the constant 1 when computing the intersection $\sd\cap \sdd$ (resp. $\sdh\cap \sddh$ in the Hermitian case). Therefore, we avoid this case in the results of this section.

\subsection{Euclidean EAQECCs}

Using the knowledge of the relative hull for two projective Reed-Muller codes, we can construct asymmetric EAQECCs. Asymmetric EAQECCs arise after noting that in quantum error-correction we consider two different types of errors, phase-shift and qudit-flip errors, which are not equally likely to occur \cite{ioffe}. Asymmetric EAQECCs have two different error correction capabilities for each of the errors, which are expressed by two minimum distances, $\delta_z$ and $\delta_x$, whose meaning is that the corresponding asymmetric EAQECC can correct up to $\lfloor(\delta_z-1)/2\rfloor$ phase-shift errors and $\lfloor(\delta_x-1)/2\rfloor$ qudit-flip errors.

Given a nonempty set $U\subset \fq^n$, we denote by $\wt(U)$ the number $\min \{\wt(v)\mid v\in U\setminus \{0\}\}$. To construct asymmetric EAQECCs, we can use the following result from \cite{galindoasymmetric}.

\begin{thm}\label{asimetricos}
Let $C_i\subset \fq^n$ be linear codes of dimension $k_i$, for $i=1,2$. Then, there is an asymmetric EAQECC with parameters $[[n,\kappa,\delta_z/\delta_x;c]]_q$, where
$$
\begin{aligned}
c&=k_1-\dim (C_1\cap C_2^\perp), \; \kappa=n-(k_1+k_2)+c, \\
\delta_z=\wt&\left(C_1^\perp\setminus \left(C_1^\perp\cap C_2 \right)\right) \text{ and }\; \delta_x=  \wt\left(C_2^\perp\setminus \left(C_2^\perp\cap C_1 \right) \right) .
\end{aligned}
$$
\end{thm}

Symmetric quantum codes can also be obtained from the previous construction by considering the minimum distance $\delta=\min \{\delta_z,\delta_x\}$ instead of the two minimum distances $\delta_z$ and $\delta_x$.

If $C_1 \subset C_2^\perp$, we have $c=0$ and in that case we do not require entanglement assistance. The asymmetric EAQECC from the previous result is called \textit{pure} if $\delta_z=\wt(C_1^\perp)$ and $\delta_x=\wt(C_2^\perp)$, and it is called \textit{impure} otherwise. For the symmetric case, the code is called pure if $\delta=\min\{\wt(C_1^\perp),\wt(C_2^\perp)\}$, and impure otherwise. 

Finding impure quantum codes is a difficult task in general. The following result supports this fact because it implies that the EAQECCs we obtain using projective Reed-Muller codes are always pure.

\begin{lem}\label{lemapuros}
Let $1\leq d_1,d_2\leq 2(q-1)$, with $d_1\not\equiv d_2 \bmod q-1$. We have that 
$$
\wt\left (\PRM_{d_1}(2)\setminus \left(\PRM_{d_1}(2)\cap \PRM_{d_2}(2)\right)\right)=\wt(\PRM_{d_1}(2))
$$
\end{lem}
\begin{proof}
If $d_2<d_1$, then $\wt(\PRM_{d_2}(2))>\wt(\PRM_{d_1}(2))$. Therefore, there is a codeword of Hamming weight $\wt(\PRM_{d_1}(2))$ in $\PRM_{d_1}(2)\setminus \PRM_{d_2}(2)$. 

On the other hand, if $d_2>d_1$, we consider two cases. If $d_1\leq q-1$, then the evaluation of the polynomial
$$
x_2\prod_{j=1}^{d_1-1}(\lambda_jx_2-x_1),
$$
where $\lambda_i\neq \lambda_j$ if $i\neq j$, $\lambda_j\in \fq^*$, has Hamming weight $q(q-d_1+1)=\wt(\PRM_{d_1}(2))$. This is easy to check using the representatives $[\fq^2\times \{1\}]\cup [\fq\times \{1\}\times \{0\}]\cup \{[1:0:0]\}$ for $\mathbb{P}^2$. This polynomial is not contained in $\sd\cap \sdd$ because this vector space is generated by $A_1^{d_1}\cup \left(\bigcup_{a_2\in Y}\{ x_1^{d_1-a_2}x_2^{a_2} \}\right)$, which does not generate the monomial $x_2^{d_1}$ that is in the support of the previous polynomial. Thus, the evaluation of this polynomial is a codeword of Hamming weight $\wt(\PRM_{d_1}(2))$ in $\PRM_{d_1}(2)\setminus \left(\PRM_{d_1}(2)\cap \PRM_{d_2}(2)\right)$. 

If $d_1\geq q$, we consider instead the polynomial
\begin{equation}\label{pol2}
x_1(x_2^{q-1}-x_1^{q-1})\prod_{j=1}^{\overline{d_1}-1}(\lambda_jx_1-x_0),
\end{equation}
where $\lambda_i\neq \lambda_j$ if $i\neq j$, $\lambda_j\in \fq^*$. As before, it is easy to check that the evaluation of this polynomial has Hamming weight $q-\overline{d_1}+1=\wt(\PRM_{d_1}(2))$. The monomial $x_1^{\overline{d_1}}x_2^{q-1}$ in the support of the previous polynomial is part of the basis from Lemma \ref{basehomogenea}. We have that $\sdd\cap \sd$ is generated in this case by $A_1^{d_1}\cup \left(\bigcup_{a_2\in Y}\{ x_1^{d_1-a_2}x_2^{a_2} \}\right)\cup \{Q_{d_1,d_2}\}$. All of these monomials, and $Q_{d_1,d_2}$, are expressed in terms of the basis from Lemma \ref{basehomogenea}. Therefore, the only way to generate a polynomial with $x_1^{\overline{d_1}}x_2^{q-1}$ in its support is to have this monomial in the expression of some element of the basis of $\sd\cap \sdd$ in terms of the basis from Lemma \ref{basehomogenea}. By checking the definitions, we see that this only happens if $\overline{d_2}=q-1$, because in that case this monomial appears in the expression of $Q_{d_1,d_2}$. However, $Q_{d_1,d_2}$ has the monomial $x_2^{d_1}$ in its support, and the polynomial from (\ref{pol2}) does not, and $x_2^{d_1}$ cannot be cancelled because no other monomial from the basis of $\sd\cap \sdd$ has this monomial in its support. Hence, the evaluation of the polynomial from (\ref{pol2}) gives a codeword of Hamming weight $\wt(\PRM_{d_1}(2))$ in $\PRM_{d_1}(2)\setminus \left(\PRM_{d_1}(2)\cap \PRM_{d_2}(2)\right)$.
\end{proof}

\begin{rem}
For $d_1\equiv d_2 \bmod q-1$, if $d_2<d_1$, then 
$$\wt\left (\PRM_{d_1}(2)\setminus \left(\PRM_{d_1}(2)\cap \PRM_{d_2}(2)\right)\right)=\wt(\PRM_{d_1}(2))$$ arguing as in the previous result. If $d_2\geq d_1$, then $$\PRM_{d_1}(2)\setminus \left(\PRM_{d_1}(2)\cap \PRM_{d_2}(2)\right)=\emptyset.$$
\end{rem}

Now we show the parameters of the asymmetric EAQECCs arising from Theorem \ref{asimetricos} when $C_1$ and $C_2$ are projective Reed-Muller codes. Note that the parameters of $\PRM_d(2)$ and $\RM_d(2)$ are in Theorems \ref{paramPRM} and \ref{paramRM}, and for $\PRM^\perp_d(2)$ we can use Theorem \ref{dualPRM}.

\begin{thm}\label{thmcuantproyasim} 
Let $1\leq d_1\leq d_2< 2(q-1)$, $d_1+d_2\not\equiv 0 \bmod q-1$, $d_1\neq q-1\neq d_2$. Let $k_1=\dim \RM_{d_1-1}(2)$ and $k_2=\dim \RM_{d_2^\perp-1}(2)$, where $d_2^\perp=2(q-1)-d_2$. Then we can construct an asymmetric EAQECC with parameters $[[n,\kappa,\delta_z/\delta_x;c]]_q$, where  $n=q^2+q+1$, $\kappa=n-(\dim \PRM_{d_1}(2)+\dim \PRM_{d_2}(2))+c$, $\delta_z=\wt(\PRM_{d_2}^\perp(2))$, $\delta_x=\wt(\PRM_{d_1}^\perp(2))$, and the value of $c$ is the following: 
\begin{enumerate}
\item If $d_1+d_2<2(q-1)$:
$$
c=\begin{cases}
 d_1+1- \min\{d_1,q-1-d_2\} &\text{ if } d_2< q-1,\\
d_1+1 &\text{ if } q\leq d_2.
\end{cases}
$$
\item If $d_1+d_2>2(q-1)$:
$$
c=\begin{cases}
k_1-k_2 +d_1+1 &\text{ if } d_1< q-1,\\
k_1-k_2+q+1-\min\{d_2^\perp,d_1-(q-1)\} &\text{ if } q\leq d_1.
\end{cases}
$$
\end{enumerate}
Moreover, this code is pure. 
\end{thm}
\begin{proof}
We consider $C_1=\PRM_{d_1}(2)$, $C_2=\PRM_{d_2}(2)$, and apply Theorem \ref{asimetricos}.
For the parameter $c$, we use Corollary \ref{cordim} with $d_1$ and $d_2^\perp=2(q-1)-d_2$, taking into account that $d_1+d_2\not \equiv 0 \bmod q-1$ implies that $d_1\not \equiv 2(q-1)-d_2\bmod q-1$, and Remark \ref{remdim}. We also note that if $d_1+d_2<2(q-1)$, then $d_1< d_2^\perp$, and if $d_1+d_2>2(q-1)$, then $d_2^\perp <d_1$. 

A direct application of Theorem \ref{asimetricos} would give us a pure quantum code with $\delta_z=\wt(\PRM_{d_1}^\perp(2))$ and $\delta_x=\wt(\PRM_{d_2}^\perp(2))$ due to Lemma \ref{lemapuros}. However, it is easy to see that if we exchange the roles of $C_1$ and $C_2$ in Theorem \ref{asimetricos}, the resulting asymmetric EAQECC has the same parameters, except that $\delta_z$ and $\delta_x$ are exchanged, which gives the result.
\end{proof}

Let $d_1\not\equiv 0\bmod q-1$, $d_2\not \equiv 0 \bmod q-1$. If $d_1+d_2=q-1$ or $d_1+d_2=2(q-1)$, we can obtain an EAQECC as in Theorem \ref{thmcuantproyasim} with $c=0$ because $\dim (\PRM_{d_1}(2)\cap \PRM_{d_2}^\perp(2))=\dim \PRM_{d_1}(2)$ by Lemma \ref{hullfacilprm}. If $d_1+d_2=3(q-1)$ or $d_1+d_2=4(q-1)$, we have $c=\dim \PRM_{d_1}(2)-\dim \PRM_{d_2}^\perp(2)$ instead, because $\dim (\PRM_{d_1}(2)\cap \PRM_{d_2}^\perp(2))=\dim \PRM_{d_2}^\perp(2)$.

\begin{ex}\label{ejasim}
We consider $\F_q=\F_9$, and we use Theorem \ref{thmcuantproyasim} with $d_1=3$, $d_2=11$. The parameters for the corresponding affine and projective Reed-Muller codes are obtained from Theorem \ref{paramRM} and Theorem \ref{paramPRM}, respectively. For the parameter $c$, we have $d_1+d_2=14<16=2(q-1)$, and $c=3+1=4$ in this case because $q=9\leq 11=d_2$. The asymmetric EAQECC that we obtain in this way has parameters $[[91,15,45/5;4]]_9$. With affine Reed-Muller codes, it is possible to obtain an asymmetric EAQECC with parameters $[[81,5,45/5;0]]_9$. If we define the rate as $\rho:=\kappa/n$, and the net rate as $\overline{\rho}:=(\kappa-c)/n$, we see that the projective code clearly has higher rate, but it also has higher net rate. 
\end{ex}

In \cite{ioffe} it is shown that the probability of phase-shift errors is between 10 and 100 times higher than the probability of qudit-flip errors, depending on the devices used for constructing qubits. Hence, it is desirable to construct EAQECCs with a higher correction capability for phase-shift errors, i.e., EAQECCs with $\delta_z\gg \delta_x$. The EAQECCs arising from Theorem \ref{thmcuantproyasim} automatically satisfy $\delta_z\geq \delta_x$. We show now how to construct codes with high asymmetry ratio $\delta_z/\delta_x$ using projective Reed-Muller codes. 

\begin{ex}\label{ejasymmetryratio}
Assume that for a certain application we want to correct 1 qudit-flip error (and detect 2), for lengths lower than 200. Therefore, we want to obtain an asymmetric EAQECC with $\delta_x=3$. If we assume that the probability of phase-shift errors is between 10 and 100 times higher than the probability of qudit-flip errors, we want to construct codes with $\delta_z$ between $30$ and $300$. If we consider the field $\F_q$, using Theorem \ref{thmcuantproyasim} it is easy to check that the asymmetric EAQECC with highest asymmetry ratio and nonzero dimension that we can obtain has parameters
$$
[[q^2+q+1,5,q(q-1)/3;2]]_q,
$$
which corresponds to $d_1=1$ and $d_2=2(q-1)-2$. By considering $q=9,11,13$, we obtain the parameters $[[91,5,72/3;2]]_9$, $[[133,5,110/3;2]]_{11}$, $[[183,5,156/3;2]]_{13}$, respectively. All of the previous codes satisfy the required conditions about the asymmetry ratio and length, and all of them surpass the quantum Gilbert Varshamov bound from \cite{matsumotoimprovedGV}. 

With affine Reed-Muller codes we can obtain instead the parameters
$$
[[q^2,3,q(q-2)/3;0]]_q.
$$
Hence, with projective Reed-Muller codes we can achieve a higher asymmetry ratio, at the expense of getting a worse net rate with respect to the affine case. We can also obtain the same asymmetry ratio as with affine Reed-Muller codes, and increase the net rate, by using projective Reed-Muller codes as we saw in Example \ref{ejasim}.
\end{ex}

It is not easy to compare the codes that we obtain with the literature because there are not many references about asymmetric EAQECCs. However, we can use the quantum Gilbert-Varshamov bound from \cite{matsumotoimprovedGV} to argue that we are obtaining quantum codes with good parameters. In Table \ref{tablaasym} we show some of the codes that we obtain that surpass the quantum Gilbert-Varshamov bound from \cite{matsumotoimprovedGV}.

{\scriptsize
\begin{table}[ht]
\caption{Codes arising from Theorem \ref{thmcuantproyasim} surpassing the quantum Gilbert-Varshamov bound from \cite{matsumotoimprovedGV}.} 
\label{tablaasym}
\centering
\begin{tabular}{||c|c|c||c|c|c|c|c||}
 \hline 
 $q$&$d_1$&$d_2$  &$n$ & $\kappa$&$\delta_x$&$\delta_z$&c \\
  \hline \hline
4&1&1&21&16&3&3&1\\
4&1&4&21&5&3&12&2\\
4&2&2&21&11&4&4&2\\
4&2&5&21&2&4&16&5\\
4&4&4&21&2&12&12&11\\
4&5&5&21&1&16&16&16\\
5&1&1&31&26&3&3&1\\
5&1&2&31&23&3&4&1\\
5&1&5&31&9&3&15&2\\
5&1&6&31&5&3&20&2\\
5&2&3&31&17&4&5&2\\
5&2&5&31&7&4&15&3\\
5&2&7&31&2&4&25&5\\
5&3&6&31&3&5&20&7\\
5&3&7&31&2&5&25&9\\
5&5&5&31&3&15&15&14\\
5&5&6&31&2&15&20&17\\
5&6&7&31&1&20&25&23\\
5&7&7&31&1&25&25&26\\
9&1&1&91&86&3&3&1 \\
9&1&2&91&83&3&4&1 \\
9&1&3&91&79&3&5&1 \\
9&1&4&91&74&3&6&1 \\
9&1&11&91&20&3&45&2 \\
9&1&12&91&14&3&54&2 \\
9&1&13&91&9&3&63&2 \\
\hline
\end{tabular}
\begin{tabular}{||c|c|c||c|c|c|c|c||}
 \hline 
 $q$&$d_1$&$d_2$  &$n$ & $\kappa$&$\delta_x$&$\delta_z$&c \\
  \hline \hline
9&1&14&91&5&3&72&2 \\
9&2&2&91&80&4&4&1 \\
9&2&3&91&76&4&5&1 \\
9&2&11&91&18&4&45&3 \\
9&2&12&91&12&4&54&3 \\
9&2&13&91&7&4&63&3 \\
9&2&15&91&2&4&81&5 \\
9&3&3&91&72&5&5&1 \\
9&3&12&91&9&5&54&4\\
9&3&14&91&3&5&72&7\\
9&3&15&91&2&5&81&9\\
9&4&13&91&4&6&63&9\\
9&4&14&91&3&6&72&12\\
9&4&15&91&2&6&81&14\\
9&11&12&91&2&45&54&57\\
9&11&14&91&1&45&72&65\\
9&11&15&91&1&45&81&68\\
9&12&13&91&1&54&63&67\\
9&12&14&91&1&54&72&71\\
9&12&15&91&1&54&81&74\\
9&13&13&91&1&63&63&72\\
9&13&14&91&1&63&72&76\\
9&13&15&91&1&63&81&79\\
9&14&14&91&1&72&72&80\\
9&14&15&91&1&72&81&83\\
9&15&15&91&1&81&81&86\\
\hline
\end{tabular}
\end{table}
}

We turn our attention now to the symmetric case. Given a symmetric quantum code obtained using the construction from Theorem \ref{asimetricos} with parameters $[[n,\kappa,\delta;c]]_q$, we can define the rate and net rate as in Example \ref{ejasim}. Fixing the length and minimum distance, if an EAQECC has better net rate than other EAQECC, while keeping the other rate constant, we will say that the first code has better parameters than the second one. In this sense, for the symmetric codes arising from Theorem \ref{thmcuantproyasim}, the following result shows that the best symmetric codes are obtained when $d_1=d_2$.

\begin{cor}\label{mejores}
We fix $1\leq d_1 < 2(q-1)$, and let $d_1\leq d_2<2(q-1)$ with $d_1\neq q-1\neq d_2$ and $d_1+d_2\not\equiv 0 \bmod q-1$. Let $k_1=\dim \RM_{d_1-1}(2)$ and $k_2=\dim \RM_{d_2^\perp-1}(2)$, where $d_2^\perp=2(q-1)-d_2$. Then the best choice for $d_2$ in Theorem \ref{thmcuantproyasim} for symmetric EAQECCs is $d_2=d_1$, which gives an EAQECC with parameters $[[n,\kappa,\delta;c]]$, where $n=q^2+q+1$, $\kappa=n-2(\dim\PRM_{d_1}(2))+c$, $\delta=\wt(\PRM_{d_1}^\perp(2))$, and 
$$
c=\begin{cases}
d_1+1-\min\{d_1,q-1-d_1\} &\text{ if } d_1< (q-1),\\
k_1-k_2 +q+1-\min\{d_1^\perp,d_1-(q-1)\}  &\text{ if } d_1>(q-1) .
\end{cases}
$$
\end{cor}
\begin{proof}
For $d_2\geq d_1$, we have that $\min\{ \wt(\PRM_{d_1}^\perp(2)),\wt(\PRM_{d_2}^\perp(2))\}=\wt(\PRM_{d_1}^\perp(2))$ from Theorem \ref{dualPRM}. Therefore, all the symmetric EAQECCs obtained from Theorem \ref{thmcuantproyasim} in this setting have the same parameter $\delta$. For $d_1=d_2$, we obtain from Theorem \ref{thmcuantproyasim} an EAQECC with the stated parameters. For $d_2> d_1$ such that $d_2\neq q-1$ and $d_2<2(q-1)$, we will see that we obtain a worse code. We have that $\dim \PRM_{d_2}(2)>\dim \PRM_{d_1}(2)$ if $d_2>d_1$, which decreases the dimension of the corresponding EAQECC with respect to the one obtained with $d_1=d_2$. From Theorem \ref{thmcuantproyasim}, we also see that $c$ is either going to increase or be the same (if $d_1+d_2\not\equiv 0 \bmod q-1$). Hence, in the sense stated before, the corresponding EAQECC with $d_2>d_1$ has worse parameters than the one obtained with $d_1=d_2$ because it has less dimension while not decreasing the parameter $c$, which gives a worse rate and net rate. 
\end{proof}

\subsection{Hermitian EAQECCs}

In this subsection we construct EAQECCs using the following Hermitian construction from \cite{galindoentanglement}. 

\begin{thm}[(Hermitian construction)]\label{hermitica}
Let $C\subset \F_{q^2}^n$ be a linear code of dimension $k$ and $C^{\perp_h}$ its Hermitian dual. Then, there is an EAQECC with parameters $[[n,\kappa,\delta;c]]_q$, where 
$$
c=k-\dim(C\cap C^{\perp_h}), \;\kappa=n-2k+c, \; \text{ and } \;\delta=\wt(C^{\perp_h}\setminus (C\cap C^{\perp_h})).
$$
\end{thm}

We see that we can use the knowledge of the Hermitian hull from Theorem \ref{hullproyectivohermitico} and Corollary \ref{dimhullherm} to compute the parameter $c$ of the EAQECCs obtained from the previous result using projective Reed-Muller codes. 

\begin{thm}\label{cuantherm}
Let $1\leq d<  q^2-1$. Then we can construct an EAQECC with parameters $[[n,\kappa,\delta;c]]_q$, where $n=q^4+q^2+1$, $\kappa=n-2(\dim \PRM_{d}(q^2,2))+c$, $\delta\geq \wt(\PRM_{d}^\perp(q^2,2))$, and the value of $c$ is given by the following: 

If $d\leq 2(q-1)$:
$$
c=
\begin{cases}
    0 &\text{ if } d \equiv 0 \bmod q-1, \\
    1 &\text{ if } 1\leq d< q-1,\\
    2 &\text{ if } q-1< d< 2(q-1).
\end{cases}
$$

If $d>2(q-1)$ and $d\equiv 0\bmod q-1$, then 
$$
c=\dim \RM_{d-1}(q^2,2) -\abs{U},
$$
and if $d>2(q-1)$, $d\not\equiv 0\bmod q-1$, then we have the upper bound
$$
c\leq \dim \RM_{d-1}(q^2,2) -\abs{U}+d-\abs{V}-\abs{W}+1.
$$
\end{thm}
\begin{proof}
We consider $C=\PRM_d(q^2,2)$ and we use the Hermitian Construction from Theorem \ref{hermitica}. We only need to prove the statements about the parameter 
$$
c= \dim \PRM_d(q^2,2)-\dim( \PRM_d(q^2,2)\cap \PRM_d^{\perp_h}(q^2,2)),
$$
for which we will use Corollary \ref{dimhullherm}. 

First, we recall that $\dim \PRM_d(q^2,2)=\abs{A_1^d}+\abs{A_2^d}+\abs{A_3^d}=\dim \RM_{d-1}(q^2,2)+d+1$ (see Remark \ref{remdim}). If $d\equiv 0 \bmod q-1$, by Theorem \ref{hullproyectivohermitico} we have that $c=\abs{A_1^d}-\abs{U}=\dim \RM_{d-1}(q^2,2)-\abs{U}$. For the case with $1\leq d\leq 2(q-1)$, we also have to consider Lemma \ref{A1hermitico}.

Now we assume $d\not \equiv 0 \bmod q-1$. If $d\leq 2(q-1)$, by Corollary \ref{dimhullherm} we would have
$$
c=\dim \PRM_d(q^2,2)-\abs{U}-\abs{V}=1+\beta_1^2.
$$ 
We have $\beta_1=0$ for $1\leq d<q-1$ and $\beta_1=1$ for $q-1<d\leq 2(q-1)$, which completes this case. 

Finally, if $d>2(q-1)$, we use Corollary \ref{dimhullherm} to finish the proof.
\end{proof}

\begin{rem}
For $2(q-1)<d\leq q^2-1$, $d\not\equiv 0 \bmod q-1$, we can write
$$
\dim \RM_{d-1}(q^2,2) -\abs{U}+d-\abs{V}-\abs{W}+1=\dim \PRM_d(q^2,2)-\abs{U}-\abs{V}-\abs{W}.
$$
We also note that $\abs{V}+\abs{W}\geq \abs{V}=\abs{T}$, but it is not necessarily equal (this would give a worse upper bound than the one given in the result). The bound given in Theorem \ref{cuantherm} for $c$ is sharp in all cases we have checked, but if we use $\abs{T}$ instead of $\abs{V}+\abs{W}$ the bound is not always sharp. We also note that we have formulas for $\abs{V}=\abs{T}$ and $\abs{U}$ in Lemma \ref{lemT} and Lemma \ref{lemacontarU} (in the Appendix), respectively.
\end{rem}

\begin{ex}
Let $q=3$. For $d=1,2,3$, using Theorem \ref{cuantherm} we obtain the parameters $[[91,85,3;1]]_3$, $[[91,79,4;0]]_3$ and $[[91,71,5;2]]_3$, respectively. All of these codes surpass the quantum Gilbert Varshamov bound from \cite{matsumotoimprovedGV}. Moreover, the code with $c=0$ surpasses the quantum Gilbert-Varshamov bound from \cite{quantumGVexigente}, which seems to be more difficult to surpass than the one from \cite{matsumotoimprovedGV} for the case $c=0$.
\end{ex}

As we stated in Remark \ref{remhullafin}, we are also able to compute the dimension of the Hermitian hull for affine Reed-Muller codes in the case $m=2$, therefore obtaining the following result.

\begin{thm}
Let $0\leq d< q^2-1$. Then we can construct an EAQECC with parameters $[[n,\kappa,\delta;c]]_q$, where $n=q^4$, $\kappa=n-2(\dim \RM_d(q^2,2))+c$, $\delta\geq \wt(\RM_d^\perp(q^2,2))$, and the value of $c$ is 
$$
c=\begin{cases}
0 & \text{ if } d< 2(q-1),\\
\dim \RM_d(q^2,2)-\abs{U_{d,d}}& \text{ if } d\geq 2(q-1),
\end{cases}
$$
where $\abs{U_{d,d}}$ is given by the expression in Remark \ref{remdimhullhermafin}. 
\end{thm}
\begin{proof}
This is a consequence of Proposition \ref{hullafin} and Proposition \ref{afinhermitico}.
\end{proof}
\begin{rem}
By Remark \ref{remdimhullhermafin}, $\abs{U_{d,d}}$ is given by the same expression as $\abs{U}$ in Lemma \ref{lemacontarU}, but considering $d=\beta_0+\beta_1 q$ instead.
\end{rem} 

\section{Appendix}

In this appendix we provide an explicit formula for $\abs{U}$, which appears in the computation of the dimension of the Hermitian hull of projective Reed-Muller codes from Corollary \ref{hullproyectivohermitico}. This formula can also be used for the dimension of the Hermitian hull of affine Reed-Muller codes (see Remark \ref{remdimhullhermafin}). 

\begin{lem}\label{lemacontarU}
Let $1\leq d <q^2-1$ with $q$-adic expansion $d-1=\beta_0 +\beta_1 q$ and let $d^\perp=2(q^2-1)-d$. We also consider the $q$-adic expansion $d^\perp=\lambda_0+\lambda_1 q + q^2$.  Then, we have
\begin{equation}\label{expU}
\abs{U}=\dim \RM_{d-1}(q^2,2)-\sum_{i=1}^4 B_i,
\end{equation}
where 
$$
B_1=\binom{q-\lambda_1-1 }{q-\lambda_1-3} \binom{\beta_1}{\beta_1-2}, B_2=\max \left\{\beta_1 \left[ \binom{q-\lambda_1-1}{q-\lambda_1-3}-\binom{q-\beta_0-1}{q-\beta_0-3} \right],0\right\},
$$
$$
B_3=\max\left\{(q-1-\lambda_1)\left[ \binom{\beta_1}{\beta_1-2}-\binom{\lambda_0+1}{\lambda_0-1} \right],0\right\},
$$
$$
B_4=\beta_1(q-1-\lambda_1)\binom{\beta_0-\lambda_1}{\beta_0-\lambda_1}\binom{\beta_1-\lambda_0-1}{\beta_1-\lambda_0-1}.
$$ 
\end{lem}
\begin{proof}
The number of monomials $x_1^{a_1}x_2^{a_2}$ such that $0\leq a_1,a_2\leq q^2-1$ and $a_1+a_2\leq d-1$ is precisely $\dim \RM_{d-1}(2)$. Now we compute the number of monomials that do not satisfy the condition $\overline{qa_1}+\overline{qa_2}\leq d^\perp-1$, i.e., the monomials such that $\overline{qa_1}+\overline{qa_2}\geq d^\perp$,and by subtracting this number from $\dim \RM_d(2)$ we obtain the cardinal of the set $U$. 

Given $x_1^{a_1}x_2^{a_2}$ with $a_1+a_2\leq d-1$ and $1\leq a_1,a_2\leq q^2-1$, we consider the $q$-adic expansions $a_1=\alpha_0+\alpha_1 q$ and $a_2=\gamma_0+\gamma_1 q$. Then we have $a_1+a_2=(\alpha_0+\gamma_0)+(\alpha_1+\gamma_1)q$. Moreover, it is easy to check that $\overline{qa_1}+\overline{qa_2}=(\alpha_1+\gamma_1)+(\alpha_0+\gamma_0)q$. However, these last expressions for $a_1+a_2$ and $\overline{qa_1}+\overline{qa_2}$ are not their $q$-adic expansions in general. We separate cases according to the different possible $q$-adic expansions for these integers, and in each case we consider the monomials $x_1^{a_1}x_2^{a_2}$ such that $a_1+a_2\leq d-1$ and $\overline{qa_1}+\overline{qa_2}\geq d^\perp$. 

\begin{enumerate}
    \item[(a)] If $\alpha_0+\gamma_0\leq q-1$ and $\alpha_1+\gamma_1\leq q-1$: we have the $q$-adic expansions $a_1+a_2=(\alpha_0+\gamma_0)+(\alpha_1+\gamma_1)q$ and $\overline{qa_1}+\overline{qa_2}=(\alpha_1+\gamma_1)+(\alpha_0+\gamma_0)q$. The condition $\overline{qa_1}+\overline{qa_2}\geq d^\perp$ cannot be satisfied in this case because $\overline{qa_1}+\overline{qa_2}<q^2$, while $d^\perp \geq q^2$ (because $d<q^2-1$). Therefore, no monomial of this type satisfies $\overline{qa_1}+\overline{qa_2}\geq d^\perp$.

    \item[(b)]  If $\alpha_0+\gamma_0\leq q-1$ and $\alpha_1+\gamma_1\geq q$: we have the $q$-adic expansion $a_1+a_2=(\alpha_0+\gamma_0)+(\alpha_1+\gamma_1-q)q+q^2$, which implies $a_1+a_2\geq q^2>d$, a contradiction with the fact that $a_1+a_2\leq d-1$. 

    \item[(c)] If $\alpha_0+\gamma_0\geq q$ and $\alpha_1+\gamma_1+1 \geq q$: we have the $q$-adic expansion $a_1+a_2=(\alpha_0+\gamma_0-q)+(\alpha_1+\gamma_1+1-q)q+q^2$, which implies that $a_1+a_2\geq q^2>d$, a contradiction. 
    
    \item[(d)] If $\alpha_0+\gamma_0\geq q$ and $\alpha_1+\gamma_1+1\leq q-1$: we have the $q$-adic expansions $a_1+a_2=(\alpha_0+\gamma_0-q)+(\alpha_1+\gamma_1+1)q$ and $\overline{qa_1}+\overline{qa_2}=(\alpha_1+\gamma_1)+(\alpha_0+\gamma_0-q)q+q^2$. In this case, we can have monomials satisfying the required conditions.
\end{enumerate}
Now we count the monomials that we consider in the case (d). The condition $a_1+a_2\leq d-1$ implies that $\alpha_1+\gamma_1+1<\beta_1$ or $\alpha_1+\gamma_1+1=\beta_1$ and $\alpha_0+\gamma_0-q\leq \beta_0$. The condition $\overline{qa_1}+\overline{qa_2}\geq d^\perp$ implies that $\alpha_0+\gamma_0-q> \lambda_1$ or $\alpha_0+\gamma_0-q= \lambda_1$ and $\alpha_1+\gamma_1\geq \lambda_0$. Hence, we have four possibilities, and we are going to compute the number of monomials satisfying each of the four possible combinations of conditions.
\begin{enumerate}
    \item If $\alpha_1+\gamma_1+1<\beta_1$, $\alpha_0+\gamma_0-q>\lambda_1$: for each value $\alpha_1\in \{0,\dots,q-1\}$, $\gamma_1$ can take any value from $\{0,\dots,\beta_1-2-\alpha_1\}$. It is not hard to check that this gives $\binom{\beta_1}{\beta_1-2}$ possible choices for the pair $(\alpha_1,\gamma_1)$. Similarly, for each value of $\alpha_0\in \{1,\dots,q-1\}$ (recall that we need to have $\alpha_0+\gamma_0\geq q$, and $\alpha_0,\gamma_0\leq q-1$), we have that $\gamma_0$ can take any value in $\{\lambda_1+q-\alpha_0+1,\dots,q-1\}$. Similarly to the previous case, this gives $\binom{q-\lambda_1-1 }{q-\lambda_1-3}$ possible choices for the pair $(\alpha_0,\gamma_0)$. Thus, we obtain
    $$
    B_1=\binom{q-\lambda_1-1 }{q-\lambda_1-3} \binom{\beta_1}{\beta_1-2}
    $$
    monomials in this case.
    \item If $\alpha_1+\gamma_1+1=\beta_1$, $\alpha_0+\gamma_0-q\leq \beta_0$, $\alpha_0+\gamma_0-q>\lambda_1$: we have $\beta_1$ possible choices for $(\alpha_1,\gamma_1)$, and for $\alpha_0,\gamma_0$ we have the condition $\lambda_1<\alpha_0+\gamma_0-q\leq \beta_0$. Note that this can only happen if $\lambda_1<\beta_0$. Assuming $\lambda_1<\beta_0$, we can compute the number of $(\alpha_0,\gamma_0)$ such that $\lambda_1<\alpha_0+\gamma_0-q$, and subtract the number of $(\alpha_0,\gamma_0)$ such that $\beta_0<\alpha_0+\gamma_0-q$. These numbers can be computed as in the previous case, and multiplying by $\beta_1$ (to take into account the possible $(\alpha_1,\gamma_1)$) we obtain
    $$
    \beta_1 \left[ \binom{q-\lambda_1-1}{q-\lambda_1-3}-\binom{q-\beta_0-1}{q-\beta_0-3} \right]
    $$
    monomials for the case $\lambda_1<\beta_0$, and 0 otherwise, which is precisely the number $B_2$ in the statement.
    
    \item If $\alpha_1+\gamma_1+1<\beta_1$, $\alpha_0+\gamma_0-q=\lambda_1$, $\alpha_1+\gamma_1\geq \lambda_0$: we can argue in the same way as the last case, taking into account that in this case we only obtain a nonzero amount of monomials if $\lambda_0<\beta_1-1$. Thus, we obtain
    $$
    B_3=\max\left\{(q-1-\lambda_1)\left[ \binom{\beta_1}{\beta_1-2}-\binom{\lambda_0+1}{\lambda_0-1} \right],0\right\}
    $$
    monomials.
    \item If $\alpha_1+\gamma_1+1=\beta_1$, $\alpha_0+\gamma_0-q\leq \beta_0$, $\alpha_0+\gamma_0-q=\lambda_1$, $\alpha_1+\gamma_1\geq \lambda_0$: in this case we obtain $\beta_1(q-1-\lambda_1)$ monomials, but only if $\beta_1-1\geq \lambda_0$ and $\lambda_1\leq \beta_0$. Therefore, there are
    $$
    B_4=\beta_1(q-1-\lambda_1)\binom{\beta_0-\lambda_1}{\beta_0-\lambda_1}\binom{\beta_1-\lambda_0-1}{\beta_1-\lambda_0-1}
    $$
    monomials of this type. Note that the product of the combinatorial numbers that appear in $B_4$ is 1 when $\beta_0-\lambda_1\geq 0$ and $\beta_1-\lambda_0-1\geq 0$, and 0 otherwise.
\end{enumerate}

Hence, the size of the set $U$ is given by 
$$
\dim \RM_{d-1}(q^2,2)-\sum_{i=1}^4 B_i.
$$
\end{proof}

\begin{rem}\label{remdimhullhermafin}
For the affine case, if $d=\beta_0+\beta_1 q$ is the $q$-adic expansion of $d$ instead of $d-1$, the proof of Lemma \ref{lemacontarU} shows that
$$
\abs{U_{d,d}}=\dim \RM_{d}(q^2,2)-\sum_{i=1}^4 B_i.
$$
This gives the dimension of the Hermitian hull for affine Reed-Muller codes in 2 variables.
\end{rem}

\section{Acknowledgements}
We would like to thank Nathan Kaplan and Jon-Lark Kim for their careful reading of this article and for pointing out a typo in a previous version of this manuscript. 

\bibliographystyle{abbrv}

\end{document}